\documentclass[12pt,preprint]{aastex}

\newcommand{\tbsp}{\rule{0pt}{18pt}}

\slugcomment{Submitted to ApJ}

\shorttitle{A deep $Chandra$ observation of NGC821}
\shortauthors{Pellegrini et al.}

\begin{document}

\title{A deep $Chandra$ look at the low L$_{\bf B}$ elliptical NGC~821:\\
X-ray binaries, a galactic wind and emission at the nucleus}

\author{S. Pellegrini\altaffilmark{1}, 
A. Baldi\altaffilmark{2}, D.W. Kim\altaffilmark{2}, G. Fabbiano\altaffilmark{2}, R. Soria\altaffilmark{2,3},
A. Siemiginowska\altaffilmark{2}, \\
M. Elvis\altaffilmark{2}}
\altaffiltext{1}{Astronomy Department, Bologna University, Italy;
silvia.pellegrini@unibo.it}
\altaffiltext{2}{Harvard-Smithsonian Center for Astrophysics, 60 Garden St, 
Cambridge, MA 02138}
\altaffiltext{3}{Mullard Space Science Laboratory, University College London, Holmbury St Mary, UK}

\begin{abstract}

The relatively nearby (distance=24.1 Mpc) elliptical galaxy NGC821,
hosting a central massive black hole but inactive at all wavelengths,
was observed with $Chandra$ for a total exposure of 230 ksec, to
search for nuclear emission and gas available for accretion.  Within
its optical image, 41 sources were detected, with spectral properties
typical of low mass X-ray binaries (LMXBs). The fractions of LMXBs in
the field and in globular clusters were determined, together with
their X-ray luminosity function (XLF) down to L(0.3--8~keV)=2$\times 
10^{37}$ erg~s$^{-1}$.  At the galactic center a source of
L(0.3--8~keV)=$6\times 10^{38}$ erg~s$^{-1}$ was detected for the
first time, slightly extended.  Its spectral shape is
quite hard ($\Gamma=1.49^{+0.14}_{-0.13}$), without intrinsic
absorption. It is surrounded by three sources with spectral
shape typical of LMXBs and luminosities on the brightest end of the
XLF. One is consistent with being pointlike; the others could be
the superposition of few point sources and/or truly diffuse emission,
with one resembling a jet-like feature.  Diffuse
emission was detected out to $R\sim 30^{\prime\prime}$, and comes
mostly from unresolved LMXBs, with a minor contribution from
other types of stellar sources. Different lines of investigation
consistently provide no evidence for hot gas. Hydrodynamical
simulations show that stellar mass losses are driven out of NGC821 in
a wind sustained by type Ia supernovae, but also hot accreting gas
within a very small inner region. A companion paper presents further
observational results from $Spitzer$ and the VLA, and
possible accretion modalities for this central massive black hole.

\end{abstract}


\section{Introduction}\label{intro}

NGC~821 is an isolated elliptical galaxy at a distance of 24.1~Mpc
(Smith et al. 2004; see Table~\ref{mainlog}), with very regular
optical isophotes of disky shape (Bender et al.  1994, Lauer et
al. 2005) and an old and metal rich stellar population as typical of
elliptical galaxies (Proctor et al. 2005). No cold (HI) or dusty ISM
has been revealed in it (Lauer et al. 2005, Sarzi et al. 2006).
NGC~821 is also one of those $\sim 30$ nearby galactic spheroids for
which a central supermassive black hole (MBH) has been claimed based
on resolved dynamical studies (Ferrarese \& Ford 2005). The mass of
this MBH is $8.5 \times 10^7 M_{\odot} $ (Table~\ref{mainlog}) and its
Eddington luminosity is $L_{Edd}\sim 1.1\times 10^{46}\, \rm
erg~s^{-1}$.  However, this MBH is extremely quiescent at all
wavelengths (see also Pellegrini et al. 2007), and represents an
excellent example of the class of "inactive" nuclei, that are very
common in the local universe (e.g., Pellegrini 2005a, Ho 2005) and key
to our understanding of the mechanisms of AGN evolution and nuclear
feedback, that have been fundamental for shaping our universe (e.g.,
Springel, Di Matteo \& Hernquist 2005).  Since it is nearby, NGC821 is
a prime target for observing a quiescent nucleus and its surroundings.

X-ray emission is a key symptom of nuclear activity resulting from
accretion onto a MBH (Rees 1984), and the hot ISM that could provide a
source of fuel is readily visible in the X-rays (e.g., Fabian \&
Canizares 1988; Loewenstein et al. 2001; Pellegrini 2005a). Previous
observations of NGC821 with the $ROSAT$ PSPC had placed an upper limit
of $2.8\times 10^{40}$ erg s$^{-1}$ (O'Sullivan et al. 2001) on the
total X-ray luminosity, suggesting a low content of hot gas (e.g., Kim
et al. 1992). With the sub-arcsecond resolution of $Chandra$ we could
attempt to resolve the different components of the X-ray emission,
i.e., the hot gas, the nucleus and the population of low mass X-ray
binaries (LMXBs; Fabbiano et al. 2004, hereafter F04).  This previous
shallow (39~ks) $Chandra$ observation, performed in 2002, revealed
diffuse emission in the central galactic region, possibly from a hot
interstellar medium.  We did not detect any source that could be
unequivocally identified as the galactic nucleus (for which we derived
$L_X/L_{Edd}<10^{-7}$), but an intriguing S-shaped feature crossing
the center of the galaxy was observed.  Explanations for this feature
included either a weak two-sided X-ray nuclear jet, or a hot gas
filament, at a temperature higher than that of the surrounding gas.
This opened the possibilities that the accretion power could end
mostly into mechanical rather than radiative power (Di Matteo et
al. 2003, Pellegrini et al. 2003a, Fabbiano et al. 2003), or that the
accretion flow may be disrupted by nuclear feedback (e.g., Ciotti \&
Ostriker 2001, Omma et al. 2004; see also Soria et al. 2006a,b).

In this paper we report the results of the analysis of deep $Chandra$
observations (for a total exposure of nearly 230 ks) aimed at
defining the properties of the various sources in the central galactic
region. With the deep $Chandra$ exposure, we can constrain all the
components (nucleus, hot gas, unresolved sources) contributing to the
emission, a necessary and critical step when the goal is to understand
the accretion process. For example, we can detect a significant
portion of the LMXB population, which was barely visible in the 2002
data, and thanks to the calculation of its X-ray Luminosity Function
(XLF) we can now discriminate between the contributions to the diffuse
emission of the unresolved LMXBs and of the hot ISM (e.g., Kim \&
Fabbiano 2004). Moreover, determining the properties of the hot ISM
also gives clues on the way the ISM evolves in this galaxy, and more
in general in galaxies of similar optical luminosity (e.g., Pellegrini
\& Ciotti 1998, Sansom et al. 2006, David et al. 2006).  A companion
paper (Pellegrini et al. 2007) presents an observational campaign
aimed at determining the properties of the nuclear emission of NGC~821
at various wavelengths, including more sensitive proprietary VLA and
$Spitzer$ IRAC observations, together with archival $HST$
observations.

The paper is organized as follows: in Section~\ref{obs} we describe
the $Chandra$ observations and data preparation; in
Section~\ref{binpop} we present the X-ray properties of the whole LMXB
population of the galaxy, also making use of $HST$ WFPC2 images
to identify those in field and in globular clusters; in
Sections~\ref{centr} and \ref{unres} we report the results of the
X-ray analysis respectively for the nuclear region and the diffuse
emission; in Section~\ref{disc} we discuss the implications of our
observations; in Section~\ref{concl} we summarize our results.

\section{$Chandra$ observations and data preparation}\label{obs}

NGC~821 was observed with $Chandra$ ACIS-S (Weisskopf et al. 2000)
seven times between November 2002 and June 2005, for a total exposure of 230
ks (Table~\ref{obslog}).  Here we consider both the archival observations
(ObsID 4408 and 4006 in Tab.~\ref{obslog}), already presented in
F04, and the new unpublished data obtained in 2004 and 2005.

The satellite telemetry is processed at the $Chandra$ X-ray Center
(CXC) with the Standard Data Processing (SDP) pipelines, to correct
for the motion of the satellite and to apply instrument
calibration. The data used in this work were reprocessed in custom mode
with the version 7.6.0 of the SDP, to take advantage of
improvements in processing software and calibration, not available at the time
of the original processing. Verification of the data products showed no
anomalies. The relative astrometry of the different observations
was corrected by using detected sources (above a 3$\sigma$ threshold) as 
reference points (http://cxc.harvard.edu/cal/ASPECT/align\_evt/).

The data products were then analyzed with the CXC CIAO v3.0.1 software
and the HEASARC XSPEC package. CIAO Data Model tools were used for
data processing, such as screening out bad pixels and producing
images in given energy bands. Calibration files in CALDB version 3.1.0
of June 2005 were used. No significant background flares were observed in
these data, so no further screening was necessary. A time-dependent
gain correction (Vikhlinin et al. 2003) was applied to the SDP Level 2
event files of the first two observations before further
analysis. This correction was already included in the SDP Level 2
event files of the other observations supplied by the CXC. The seven
observations were then coadded to produce a deep image, using the CIAO
task $merge\_all$, which takes as input the event files and the
relative aspect solution (ASOL) files, and reprojects the coordinates
of all the observations onto the first one.

Below we report the results of the analysis of these data.

\section{The Low Mass X-ray Binaries population}\label{binpop}

\begin{figure}
\epsscale{.9}
\plotone{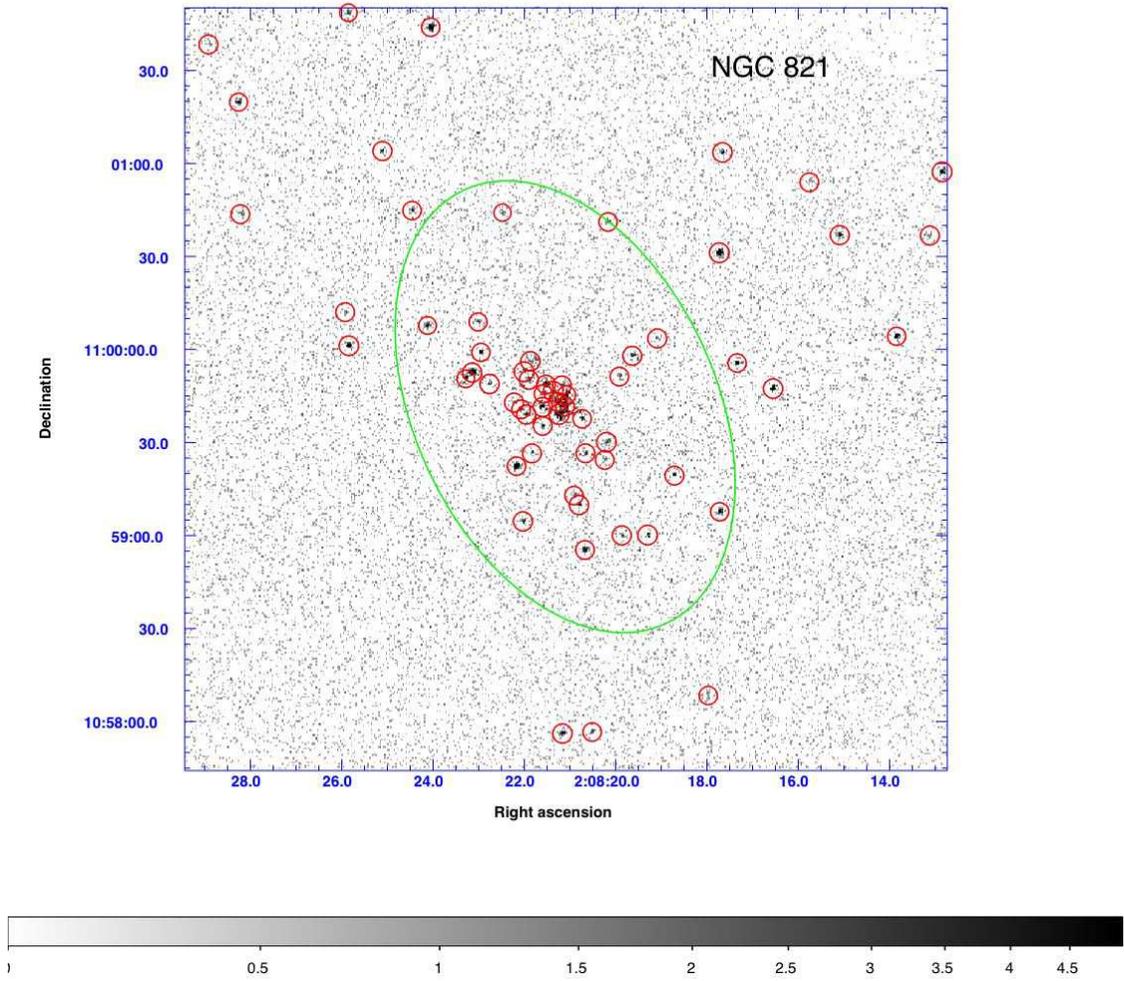}
\caption{Merged ACIS-S image of NGC821, with the D25 ellipse of
Table~\ref{mainlog} overplotted in green and the X-ray sources
detected by $wavdetect$ marked in red with circles of
$3^{\prime\prime}$ radii (Sect.~\ref{binpop}).}\label{Fig1} 
\end{figure}

Figure~\ref{Fig1} shows the merged X-ray image, with the D25 ellipse
of NGC821 overplotted (this ellipse is the isophote of the 25.0
B-mag/square arcsec brightness level, Tab.~\ref{mainlog}); also marked
are the X-ray sources detected by the CIAO task $wavdetect$.  Of the
104 sources detected within the ACIS-S3 CCD, 41 lie inside the D25
ellipse. Except for minor contamination from interlopers (see
Sect.~\ref{xlf} and \ref{stars}), and four particular sources in the
central region (discussed in Sect.~\ref{centr} below), these sources are
pointlike and represent the LMXBs population of NGC821 (see Fabbiano
2006 for a review on X-ray binary populations of galaxies). 

For the pointlike sources within D25, counts were extracted within a circle centered on
the $wavdetect$-determined source position, and background counts were
estimated locally in an annulus surrounding the source. We chose the
source extraction radius to be the 95\% encircled energy radius at 1.5
keV (that is varying as a function of the off-axis angle), with a
minimum of $3^{\prime\prime}$ near the aim point. Similarly, the
background was estimated for each source from an annulus surrounding it,
with inner and outer radii of 2 and 5 times the source radius respectively. 
When nearby sources are found within the background
region, they are excluded before measuring the background counts. Net
count rates were then calculated with the effective exposure (including
vignetting) for both the source and background regions. Errors on counts
were derived following Gehrels (1986).  When the source extraction regions
of nearby sources overlap, to avoid an overestimate of
their source count rates, we calculated the source counts from a
pie-sector, excluding the nearby source region, and then rescaled
them based on the area ratio of the chosen pie to the full disk. Once
the correction factor is determined, the same factor can be applied to
correct counts in all energy bands. For a small number of sources
which overlap with nearby sources in a more complex way (e.g., overlap
with more than 2 sources), instead of correcting the aperture
photometry, we used the source cell determined by $wavdetect$ to extract
the source counts.

The range of net detected counts for the pointlike sources is $\sim
8-300$, corresponding to 0.3--8 keV luminosities of $2\times 10^{37}$
-- $8\times 10^{38}$ erg s$^{-1}$, when using the energy conversion
factor described in Sect.~\ref{xlf} below. Most of the sources are too
faint for a detailed spectral analysis, therefore their hardness ratio
and X-ray colors were calculated in order to characterize their
spectral properties.  The X-ray hardness ratio is defined as HR =
(H$-$S) / (H$+$S), where S and H are the net counts in the $0.5 - 2.0$
keV and $2.0 - 8.0$ keV bands respectively.  Following the
prescription of Kim et al. (2004), the X-ray colors are defined as C21
= log(C1/C2) and C32 = log(C2/C3), where C1, C2 and C3 are the net
counts respectively in the energy bands of 0.3--0.9 keV, 0.9--2.5 keV and
2.5--8 keV. These counts were corrected for the temporal QE variation,
referring them all to the first observing epoch (Nov. 2002,
Tab.~\ref{obslog}), and for the effect of the Galactic absorption
(using the value in Table~\ref{mainlog}).  By definition, as the X-ray
spectra become harder, the HR increases and the X-ray colors decrease.
For faint sources with a small number of counts, the formal
calculation of the HR and colors often results in unreliable errors,
because of negative net counts in one band and an asymmetric Poisson
distribution. Therefore, we applied a Bayesian approach developed by
Park et al. (2006), which models the detected counts as a
non-homogeneous Poisson process, to derive the uncertainties
associated with the HR and colors.

The main properties of the point sources detected in the merged
observation within the D25 ellipse are summarized in Table~\ref{kim},
where the sources are listed in order of increasing distance from the
galactic center.  Column (1) gives the IAU name (following the
convention that the name should be "CXOU Jhhmmss.s$+/-$ddmmss");
columns (2) and (3) give the J2000.0 source position and column (4) its
associated uncertainty at the 95\% confidence level, determined based
on the source net counts and off-axis angle, from the empirical
formula given by Kim et al. (2007); column (5) gives the projected
distance $R$ from the galactic center; columns (6) and (7) give the
net counts and the $1\sigma$ error; columns (8) and (9) give the flux
and luminosity in the 0.3--8 keV band, derived using the energy
conversion factor described in Sect.~\ref{xlf} below; columns
(10)--(15) give the hardness ratio HR, the C21 and C32 colors defined
above, together with the respective error determined at a $1\sigma $
significance level.

\subsection{Spectral properties and optical identification}\label{kimkun}

\begin{figure*}
\hskip -1.1truecm
\epsscale{1.13}
\plottwo{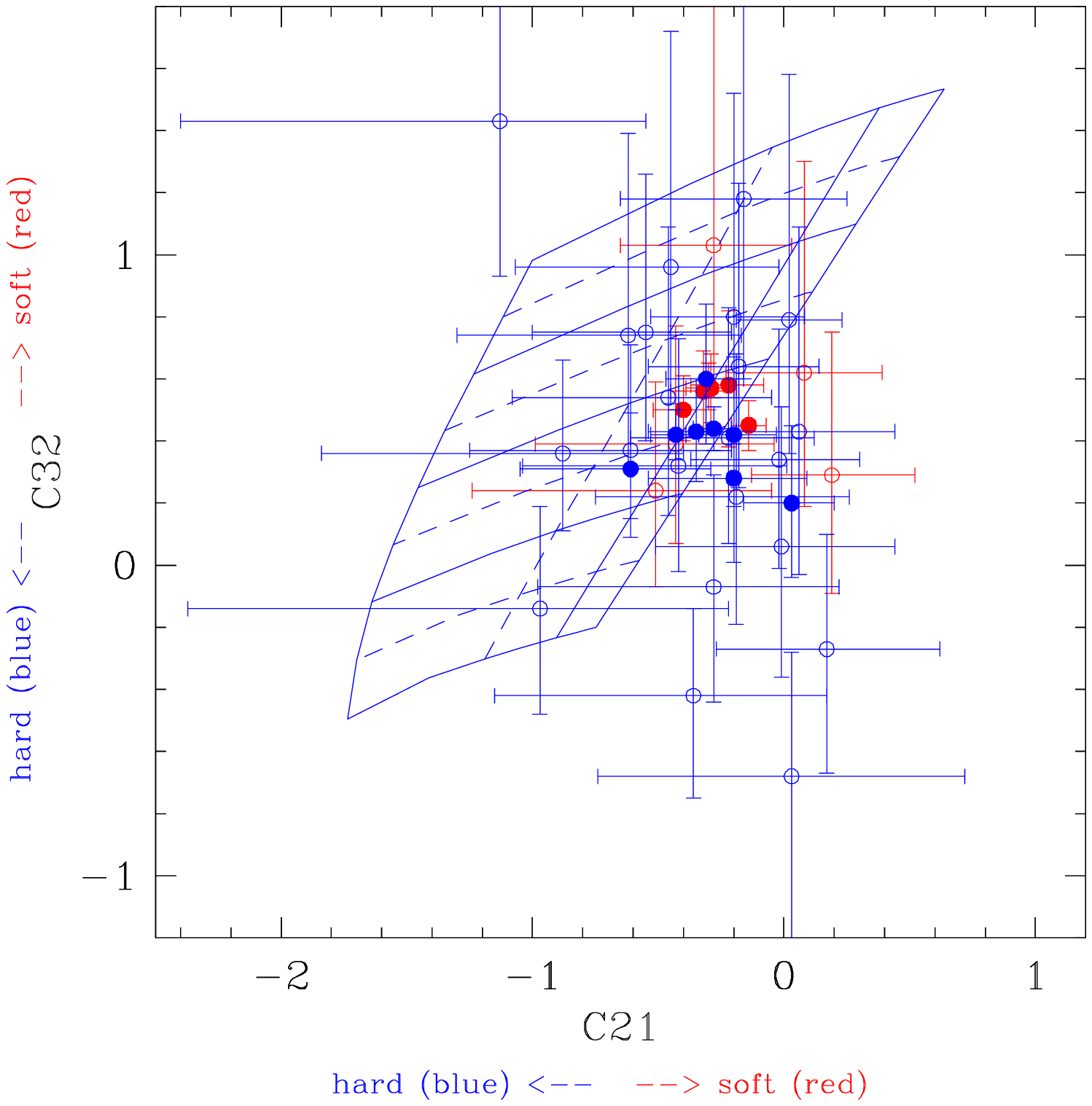}{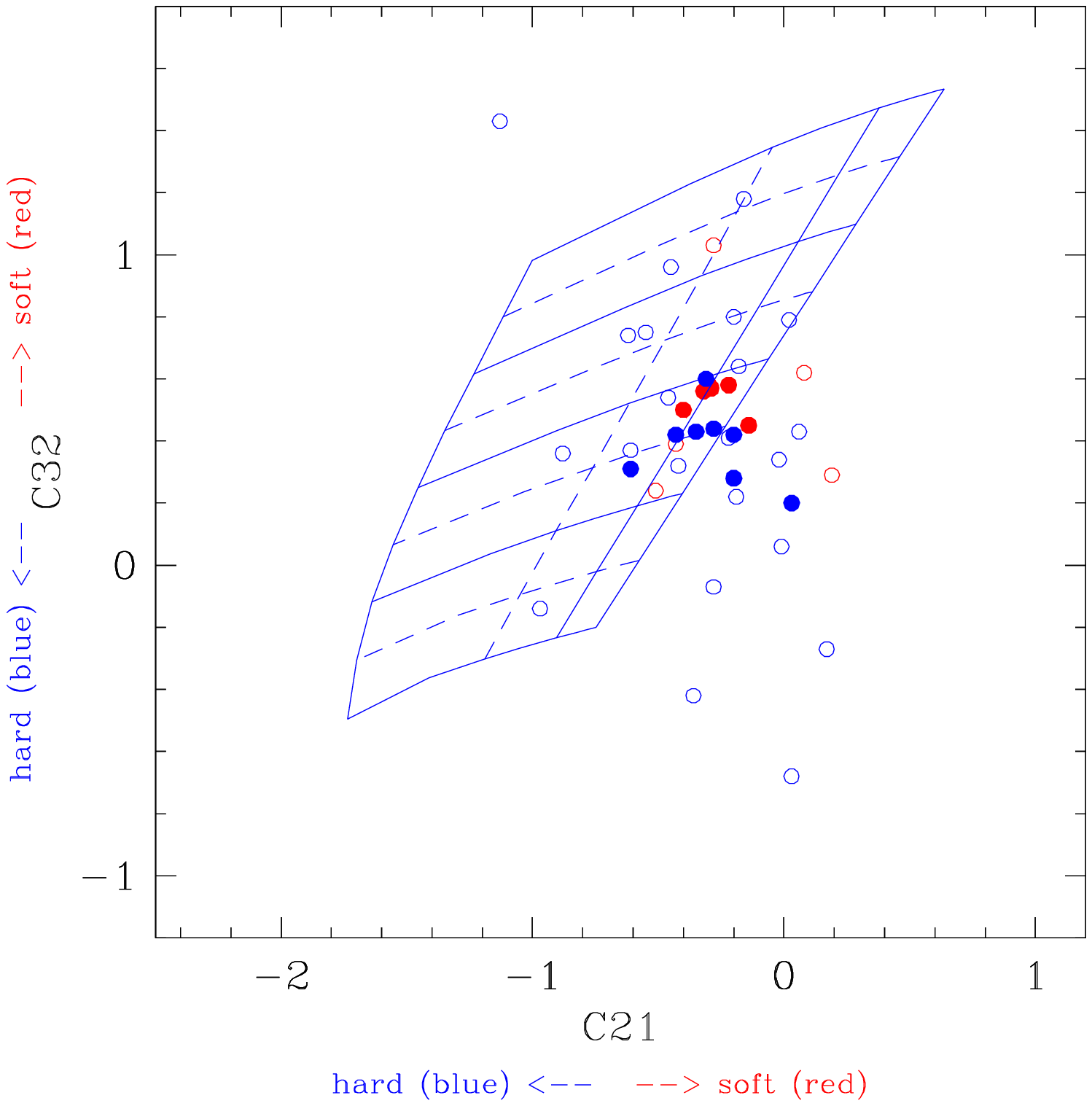}
\vskip -2truecm
\caption{X-ray colors of point sources, left: with errorbars, and right:
without errorbars (Sect.~\ref{kimkun}). The grid indicates photon index
values of 0, 1, 2, 3, 4 (from bottom to top) and $N_H$ values (from
right to left) of $10^{20}, 10^{21}, 5\times 10^{21}$ (dashed one),
$10^{22}$ cm$^{-2}$.  Red circles correspond to sources at $R< 10"$,
blue circles to sources between $R=10^{\prime\prime}$ and the ellipse 
D25.
Filled circles are sources with net counts $> 30$, open circles with
net counts $<$ 30.}\label{colors}
\end{figure*}

Figure~\ref{colors} shows the C32 -- C21 distribution of the point source
population; the grid gives
the location of spectra described by power laws with given values of
the photon index $\Gamma$ and intrinsic column density $N_H$.  In
agreement with previous spectral studies of LMXB populations (e.g.,
Kim, Fabbiano \& Trinchieri 1992; Irwin, Athey \& Bregman 2003,
Fabbiano 2006), most
well-defined colors fall near a typical $\Gamma=1.5-2.0$ spectrum,
with no intrinsic absorption.

\begin{figure}[ht]
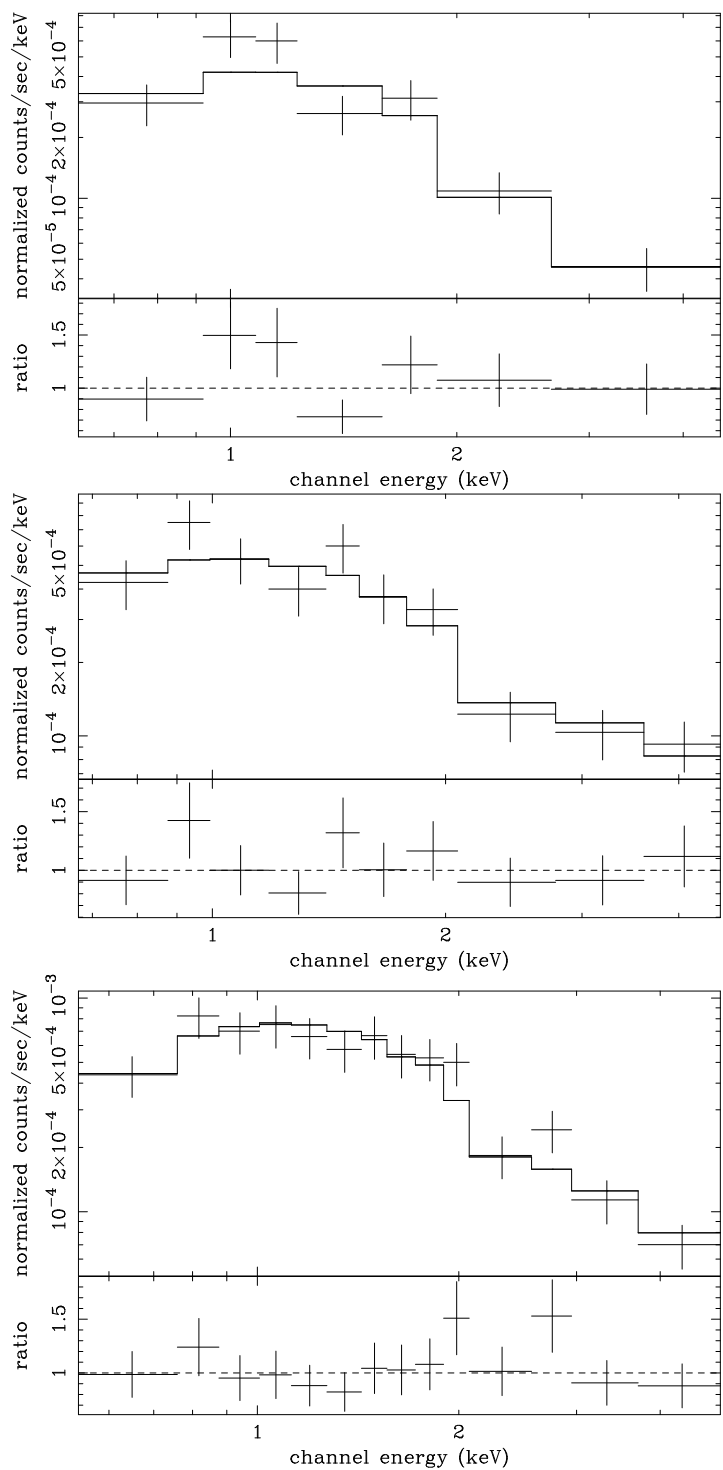

\centering
\includegraphics[scale=.4,angle=-90]{f3a.eps}\\
\includegraphics[scale=.4,angle=-90]{f3b.eps}\\
\includegraphics[scale=.4,angle=-90]{f3c.eps}
\caption{The $Chandra$ ACIS-S spectra of the brightest point sources in Tab.~\ref{kim}
together with their best fit power law models (Tab.~\ref{baldi}).
From top to bottom: J020821.5$+$105948, J020822.2$+$105922 and  J020823.1$+$105952  
(see also Sect.~\ref{kimkun}).}\label{poisp} 
\end{figure}

Three of the sources in Tab.~\ref{kim} have been detected with a
number of net counts sufficiently high for a specific spectral
analysis (they are marked in Tab.~\ref{kim}). We performed such
analysis, extracting from the merged event file the spectra for these
sources, using an extraction radius of $2^{\prime\prime}$; this radius
was chosen to avoid contamination from nearby fainter
sources. Instrumental response files, weighted by the exposure time of
each individual observation, were also generated; the background
spectrum was derived from a source-free region outside the optical
body of the galaxy. The results of the spectral analysis
(Tab.~\ref{baldi}, Fig.~\ref{poisp}) confirm that these sources are
well described by power laws with low-to-moderate intrinsic
absorption, as typical for the LMXB population. Their intrinsic 0.3--8 keV
luminosities are 4.1, 6.9 and 8.9 $\times 10^{38}$ erg s$^{-1}$.

With the aid of $HST$ WFPC2 images, we then investigated possible
associations of our detected point sources with LMXBs belonging to the
galactic field or to globular clusters (GC).  Kundu \& Whitmore (2001)
studied the globular cluster population of NGC~821 using $HST$ WFPC2
images in the $V$ (F555W) and $I$ (F814W) filters; these optical
images cover a large fraction ($\sim 60$\%) of the D25 ellipse of
NGC821.  Kundu \& Whitmore (2001) identified 105 GC candidates, whose
positions, magnitudes and colors were kindly provided to us by
A. Kundu (2007, private communication). This list of GCs was
cross-correlated with the X-ray source list of Tab.~\ref{kim}, in
order to look for GC/X-ray source coincidences.  After correction for
a small shift of the optical image (see Sect.~\ref{nucl} for more
details), 6 matches were found well within the 95\% uncertainty radius
of the X-ray position; another potential match has quite a distorted
shape and uncertain magnitude in the optical (CXOU
J020817.7$+$105907), since it is located at a chip edge, and another
one has an offset of 1.24 times the X-ray positional
uncertainty. These 8 sources are marked in Tab.~\ref{kim} (see also
Fig.~\ref{hstimage2}). From a $V$--$(V-I)$ diagram of the whole sample
of 105 GC candidates, one can see that LMXBs are preferentially found
in the brightest clusters (4 matches reside in the 5 brightest GCs,
and 6 in the 12 brightest ones), following a general trend already
reported (e.g., Sarazin et al. 2003, Kim E. et al. 2006) that is
explained as the probability of hosting a LMXB increasing with the GC
luminosity. The LMXBs of NGC821 also tend to be associated with the
redder GCs, in agreement with previous works (e.g., Kundu et
al. 2007), even though the small number of sources here does not make
this result statistically significant.

From the $HST$ WFPC2 images above, in the overlap region with the D25
ellipse, 17 X-ray sources have no optical counterpart, at a 50\%
completeness limit of 24.1 V-mags (A. Kundu 2007, private
communication\footnote{This V-magnitude value together with those
quoted in Sect.~\ref{nucl} have been computed by Kundu \& Whitmore
(2001) assuming the Burstein \& Heiles (1982) reddening.}); these can
be considered field LMXB candidates, and are marked in Tab.~\ref{kim}.
In the following we further limit the analysis to sources located at
radii larger than $10^{\prime\prime}$, to avoid possible completeness
problems near the galactic center (see Sect.~\ref{xlf} below). Then,
it turns out that $\sim 30$\% of the X-ray sources lie in GCs.  From
$Chandra$ studies this fraction is known to increase with the GC
specific frequency of the galaxy, or the morphological type (e.g.,
Sarazin et al. 2003), and goes from $\sim 20$\% in S0s to 30--50\% in
Es (e.g., Fabbiano 2006 and references therein). The fraction
estimated here for NGC821 is therefore close to what expected, given
also its morphological type (E6).  The fraction of GCs that host an
X-ray source is $\sim 6.6$\%, a value broadly consistent with what
typically observed previously from $Chandra$ studies (3--5\%, Sarazin
et al. 2003, Kim, E. et al. 2006; see also Kundu et al. 2007), but
somewhat higher, likely because of the larger depth of our X-ray
observation compared to the previous analyses.  Finally, there is no
statistically significant difference between GC and field LMXBs.  The
present sample of X-ray sources is too small to investigate other
statistical properties, such as the X-ray spectral difference 
between LMXBs in red and blue GCs or the role of galactocentric distance.

\subsection{The LMXB X-ray Luminosity Function}\label{xlf}

\begin{figure}[ht]
\vskip -2truecm
\epsscale{.9}
\hskip -2.05truecm
\plotone{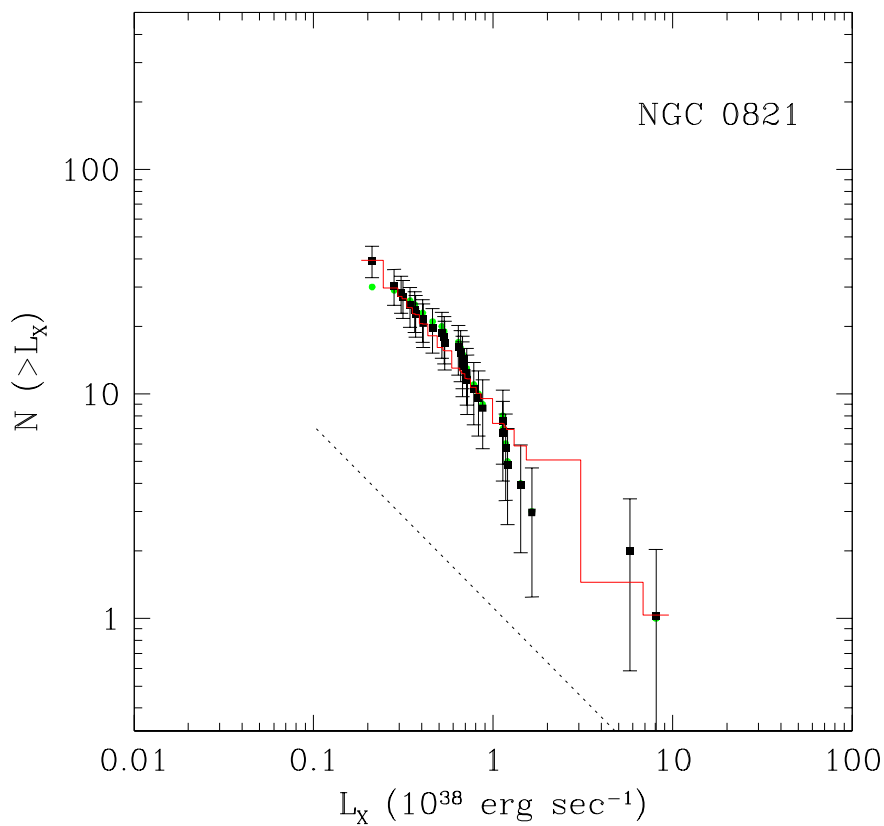}
\vskip -3.8truecm
\caption{Cumulative XLF of point sources detected in the merged ACIS-S image 
within the D25 ellipse, with a central circle of $10^{\prime\prime}$ radius
excluded (Sect.~\ref{xlf}).
Black squares with error bar show the bias-corrected function; green
circles are the uncorrected function; the red histogram gives the best fit
single power-law; the dotted line shows the expected number of cosmic
X-ray background sources from the logN--logS relation of ChaMP plus
CDF data (Kim et al. 2007).}\label{Fig2} 
\end{figure}

For the point sources falling within the D25 ellipse we estimated an
XLF, that in differential form can be expressed as: 

\begin{equation} {dN\over dL_X} = k \left( {L_X\over 10^{38}\,{\rm
erg\, s^{-1}}} \right) ^{-\beta}.  \end{equation} \noindent To
construct the XLF, we calculated fluxes and luminosities of the point
sources in the 0.3--8 keV band with an energy conversion factor (ECF)
corresponding to an assumed power law spectral shape with $\Gamma
=1.7$ and Galactic $N_H$ (see Fig.~\ref{colors} for a justification of
this assumption). The ECF was calculated with the arf (auxiliary
response file) and the rmf (redistribution matrix file) generated for
each source in each observation. As done for the colors, we took
into account the temporal QE variation by calculating the ECF in each
observation and then taking an exposure-weighted mean ECF. The ECF
over the 0.3--8 keV band varied by $\sim 2.5$\% between 2002 and 2005.
We then corrected the XLF to eliminate biases and incompleteness
effects, which could affect the lowest luminosity range, causing an
artificial break (Kim \& Fabbiano 2003, 2004; Kim et al. 2006).  Our
procedure for correcting XLFs derived from $Chandra$ data was
developed and first applied to NGC1316 by Kim \& Fabbiano (2003; see
the Appendix therein, where the method is discussed in detail), and
was then applied to a sample of early-type galaxies by Kim \& Fabbiano
(2004). A similar procedure had also been used to correct the XLF of the
Antennae galaxies (Zezas \& Fabbiano 2002). In brief, we simulated
$\sim 20,000$ point sources, added them one by one to the observed
image, and then run $wavdetect$ to determine whether the added source
was detected.  With this procedure we correct simultaneously for
incompleteness near the detection threshold, Eddington bias (Eddington
1913) and source confusion.  In the simulations, we assumed a typical
XLF of differential slope of $\beta =2$; however, the adopted slope
does not affect the results significantly, and neither does the
assumed radial distribution of point sources (which we assume to
follow the optical light, as observed in many elliptical galaxies, see
Fabbiano 2006).  We excluded central sources with
$R<10^{\prime\prime}$ in Tab.~\ref{kim}, because of the large
photometric error caused by confusion with other overlapping sources,
and by the possible presence of some diffuse emission; these
conditions make the incompleteness corrections uncertain.  With the
above procedure, we established the 90\% completeness limit (i.e., the
luminosity at which 10\% of sources would not be detected inside the
D25 ellipse, excluding the central 10$^{\prime\prime}$) to be
$L_X=3\times 10^{37}$ erg s$^{-1}$ (or $F_X=4.3\times 10^{-16}$ erg
cm$^{-2}$ s$^{-1}$) in the 0.3--8 keV band.  We can reliably correct
the XLF to a luminosity of $\sim 2/3$ of the 90\% limit. The resulting
cumulative XLF [i.e., the number of sources $N(>L_X)$] is shown in
Fig.~\ref{Fig2}.  From the log N -- log S relation derived for the
ChaMP plus CDF data (Kim et al. 2007), we then estimated that the
number of expected cosmic background sources (mostly AGN) falling
within the D25 ellipse and brighter than the 90\% completeness limit
is 3, corresponding to a source contamination of 7\%.

To establish the functional form of the XLF, we fitted the
bias-corrected differential XLF [eq. (1)] with single power-laws,
using both the Cash and $\chi^2$ statistics in the CIAO $Sherpa$
modeling and fitting application. We found that a single power-law
represents well the XLF, with a best fit slope $\beta =2.0\pm 0.3$,
compatible with other power-law fits of LMXB XLFs (Kim \& Fabbiano
2004; Kim et al. 2006) and a best fit amplitude $k=8.4_{-2.5}^{+3.1}$.
Even without the correction for incompleteness, that is significant
only at the point corresponding to the faintest fluxes ($L_X=2\times
10^{37}$ erg s$^{-1}$), a single power law is a good fit all the way
down to $L_X =3\times 10^{37}$ erg s$^{-1}$. Figure~\ref{Fig2} may
suggest a low luminosity break around $L_X\sim 5\times 10^{37}$ erg
s$^{-1}$. This break was suggested to be a universal feature of LMXB
populations (Gilfanov 2004), but is not observed in all cases (Kim
et al. 2006). For NGC821 there is no compelling evidence for
such a break, given the size of the error bars, and the good
fit with a single power-law model.  The combined LMXB XLF (Kim \&
Fabbiano 2004; Gilfanov 2004) shows a high luminosity break at
$L_X>(5.0\pm 1.6)\times 10^{38}$ erg s$^{-1}$, that may originate
from the presence of neutron star and black hole binary
populations. We cannot constrain here the high luminosity shape of the
XLF because of the small number of luminous sources (only 2
sources have $L_X > 5 \times 10^{38}$ erg s$^{-1}$).

In contrast to the uniform slope, the amplitude of the XLF varies
widely from galaxy to galaxy, reflecting the varying content of LMXBs
per unit galactic luminosity. For their sample of early type galaxies,
from integration of the XLF above $L_{X,min}=10^{37}$ erg s$^{-1}$,
Kim \& Fabbiano (2004) estimate the total $L_X$ due to LMXBs and the
average ratio $L_X/L_B=(0.9\pm 0.5)\times 10^{30}$ erg s$^{-1}$
$L_{B,\odot}^{-1}$. For NGC821 this ratio is $0.4\times 10^{30}$ erg
s$^{-1}$ $L_{B,\odot}^{-1}$ (see Sect.~\ref{xlfunres}), lower than the
mean value by $1\sigma$. This ratio is known to correlate with the
total specific frequency of globular clusters, but this quantity
unfortunately is not available for NGC821. Based on the $L_X/L_B$
ratio, we would expect NGC821 to have a relatively poor GC population.

\section{The central region of NGC821 -- resolved sources}\label{centr}

\begin{figure}[ht]
\epsscale{.9}
\plotone{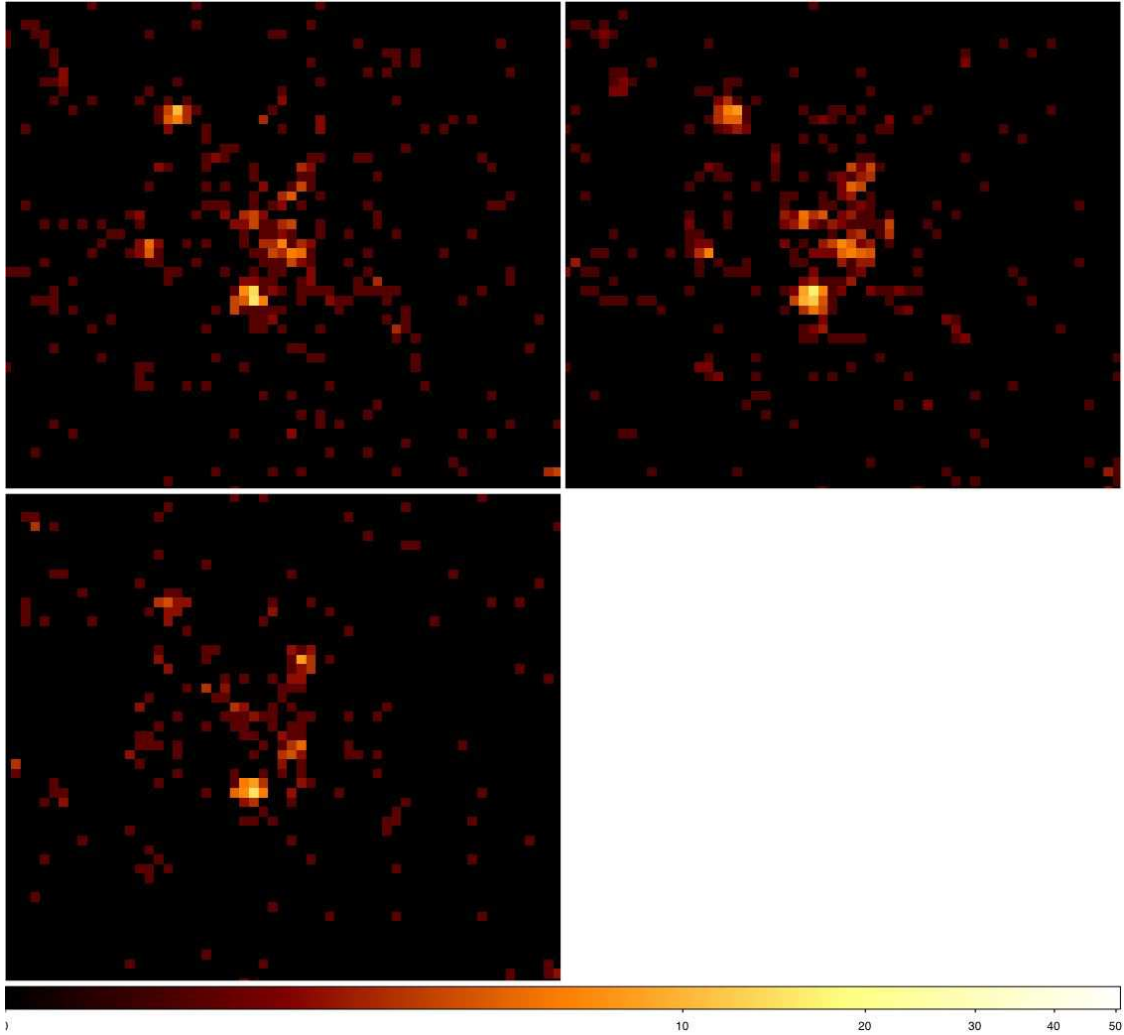}
\caption{Co-added $Chandra$ ACIS-S image of a central field of $25^{\prime\prime}
\times 25^{\prime\prime}$
in three bands: 0.3--1 keV (top left), 1--2 keV (top right), and 2--4 keV
(bottom left). North is up and east is to the left.
The bar below the figure gives the correspondance between the
color scale and the number of counts/pixel (the size of 1 px is $0\farcs5$). 
See Sect.~\ref{centr}
for more details.}\label{3bands}
\end{figure}

\begin{figure}[ht]
\epsscale{.9}
\plotone{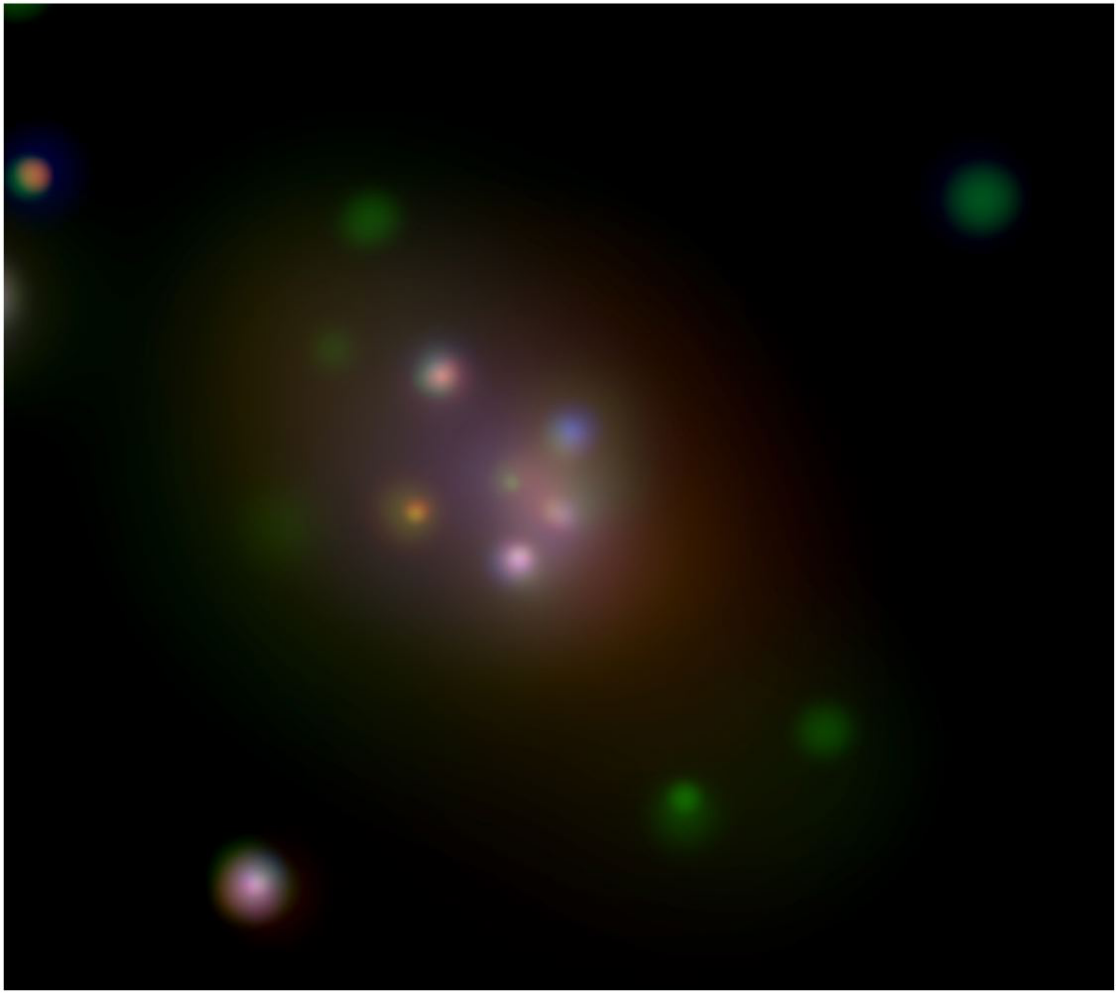}
\caption{Co-added smoothed "true-color" image of a central $55^{\prime\prime} 
\times 55^{\prime\prime}$ field  (see details in Sect.~\ref{centr}).
The red, green and blue
colors correspond to the 0.3--1 keV, 1--2 keV and 2--4 keV bands
respectively. North is up and east is to the left.
The diffuse emission follows the optical shape of the galaxy (see
Tab.~\ref{mainlog}, Fig.~\ref{Fig1}
and Fig.~\ref{hstimage2} below).}\label{truecol}
\end{figure}

Figure~\ref{3bands} shows the central region of the image obtained
from the merged data, in three spectral bands (0.3--1 keV, 1--2 keV
and 2--4 keV) that cover the energy range where most of the counts are
detected.  From these three images in different spectral bands we also
obtained an adaptively smoothed (using the CIAO task `csmooth') color
image (Fig.~\ref{truecol}), where the data were smoothed using scales
ranging from 1 to 20 pixels ($0\farcs5-10^{\prime\prime}$).  Given the
small extent of the region considered, no exposure correction was
needed.  Figure~\ref{truecol} shows diffuse emission at the center,
that follows the optical shape of the galaxy: it is elongated in the
NW-SE direction, roughly aligned with the optical major axis (see also
Tab.~\ref{mainlog}, Fig.~\ref{Fig1} and Fig.~\ref{hstimage2} below).
Most of the diffuse emission is concentrated well within one $R_e$
(that is $43\farcs9$, while the size of Fig.~\ref{truecol} is
$55^{\prime\prime}$). Its color indicates emission mostly in the soft
(red) and medium (green) energy bands.  Harder emission from a few
central point-like sources is also visible.

Below we concentrate on the X-ray analysis of the sources detected in
the central region; in Sect.~\ref{unres} we report the analysis of the
diffuse emission, for the central region and the whole galaxy.

\subsection{Detection and spatial properties \label{match}}

\begin{figure}[ht]
\epsscale{.9}
\plotone{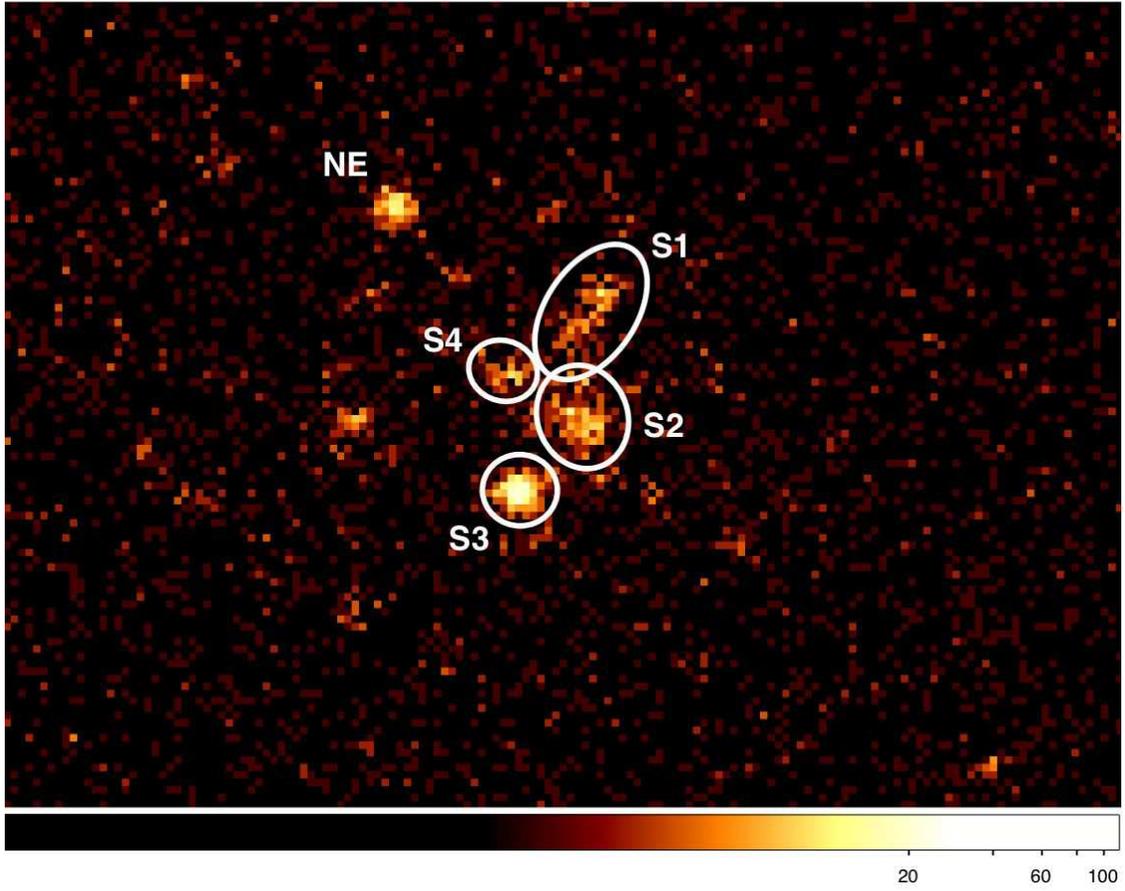}
\caption{
The ACIS-S image in the 0.3--6 keV band of the central $40^{\prime\prime}\times 25^{\prime\prime}$
region of NGC~821, with the sources and their extent
detected by the CIAO task $wavdetect$ (Sect.~\ref{match}).
North is up and east is to the left.
Here 1 px=$0\farcs25$. The semi-major
and semi-minor axes for S1--S4 are respectively $2\farcs5\times 1\farcs5$,
$1\farcs7\times 1\farcs5$, $1\farcs2\times 1\farcs2$ and $1\farcs2\times 
1\farcs0$.}\label{s1-s4}
\end{figure}

We can now establish with higher significance than in F04 the
morphology of the central X-ray emission.  In particular, F04 had
identified three possibly connected extended central emission regions
(that they called S1 through S3) and a point-like source to the
north-east of this grouping (labelled NE source).  $wavdetect$ detects
all these emission regions in the merged data, plus 
another source that we call S4 (Fig.~\ref{s1-s4}). F04 had found
the spatial distribution of photons of the S1--S3 sources to be more extended
than expected for point-like sources seen through the $Chandra$
mirrors, and suggested that they may be part of an S-shaped jet or
filament, possibly connected with low-level or past nuclear
activity. With the longer exposure, it is apparent (Fig.~\ref{s1-s4})
that the emission of S3 may be point-like, suggesting that this source
is a luminous LMXB, but the other sources are more extended.

In order to establish the spatial properties of S1--S4, following F04
we compared the spatial distribution of their counts with that of the
on-axis image of the quasar GB~1508+5714, a hard point-like source
(Siemiginowska et al. 2003). GB~1508+5714 gives a good representation
of the $Chandra$ ACIS-S PSF for our analysis, since it lies at the
same distance from the aimpoint as S1--S4 and is similarly hard (see
Sect.~\ref{specnucl} below).  The image of GB~1508+5714 contains 5,300
counts within 2$^{\prime\prime}$ of the centroid of the count
distribution; the ratio of counts within the
1$^{\prime\prime}$--2$^{\prime\prime}$ annulus to those in the central
circle of 1$^{\prime\prime}$ radius is $Ratio(PSF)$=0.043$\pm0.001$
(1$\sigma$).  The isolated NE source in this central field
(Fig.~\ref{s1-s4}) has a total of 173 source counts and
$Ratio(NE)$=0.123$\pm$0.083, a value consistent with that of our
reference quasar, within the 1$\sigma$ error.

The total number of source counts for S1--S4 are 184, 250, 326 and
84 respectively; the analogous ratios for the background-subtracted
counts, around the centroids determined by $wavdetect$, are $Ratio(S1)$=
0.59$\pm$0.11, $Ratio(S2)$=0.77$\pm$0.10, $Ratio(S3)$=0.20$\pm$0.06, and 
$Ratio(S4)$=0.80$\pm$0.15.  Therefore the
spatial distribution of the counts from S1--S4 is extended,
when we compare it with that of the quasar. If, more conservatively, we use
as a benchmark the NE source, we find that S3 is point-like, while S1,
S2 and S4 are all significantly more extended than the NE source (at 3.5,
4.6 and 4$\sigma$ respectively).  Although the significance of the NE
source is such that its $Ratio$ is less well determined than that of
the quasar, this source would be affected by similar small amounts of
smearing in the merging process as the S1--S4 sources. [Note that in
F04 the NE source had a $Ratio(NE)$=0.057$\pm0.054$,
consistent within the errors with the value we find here,
although the nominal value is smaller; this justifies using NE as a
benchmark.]

In conclusion, excluding source S3, the counts of S1, S2 and S4 cannot
come entirely from a point source. While the previous $Chandra$
observation was suggestive of a central emission feature
intrinsically elongated, possibly mostly due to diffuse emission (F04), the
present imaging analysis suggests a cluster of extended sources.  With
these data it remains possible that each source is truly
extended or produced by several point-like components, or
alternatively due to a point-like component embedded in truly diffuse
emission.

\subsection{Position of the nucleus - S2}\label{nucl}

From the Two Micron All Sky Survey (2MASS, Skrutskie et al. 2006), the
center of NGC821 is at RA=$02^h$ $08^m$ $21^s$\hskip-0.1truecm.14,
Dec=$+10^{\circ}$ $59^\prime$ $41\farcs7$ (J2000), with an uncertainty
of $1\farcs25$ (at 95\% confidence), as reported in the NASA/IPAC
Extragalactic Database (NED). Considering also the $Chandra$ aspect
uncertainty\footnote{The $Chandra$ absolute astrometric accuracy is discussed
at http://cxc.harvard.edu/cal/ASPECT/celmon/} (the $90\%$ uncertainty circle
has a radius of $\la
0\farcs6$, and the $68\%$ circle of $\la 0\farcs25$), 
this position agrees with that of the extended source S2
determined by $wavdetect$, that is RA=$02^h$ $08^m$
$21^s$\hskip-0.1truecm.10, Dec=$+10^{\circ}$ $59^\prime$ $41\farcs6$,
with an offset of just $0\farcs6$.

\begin{figure}[ht]
\epsscale{.9}
\plotone{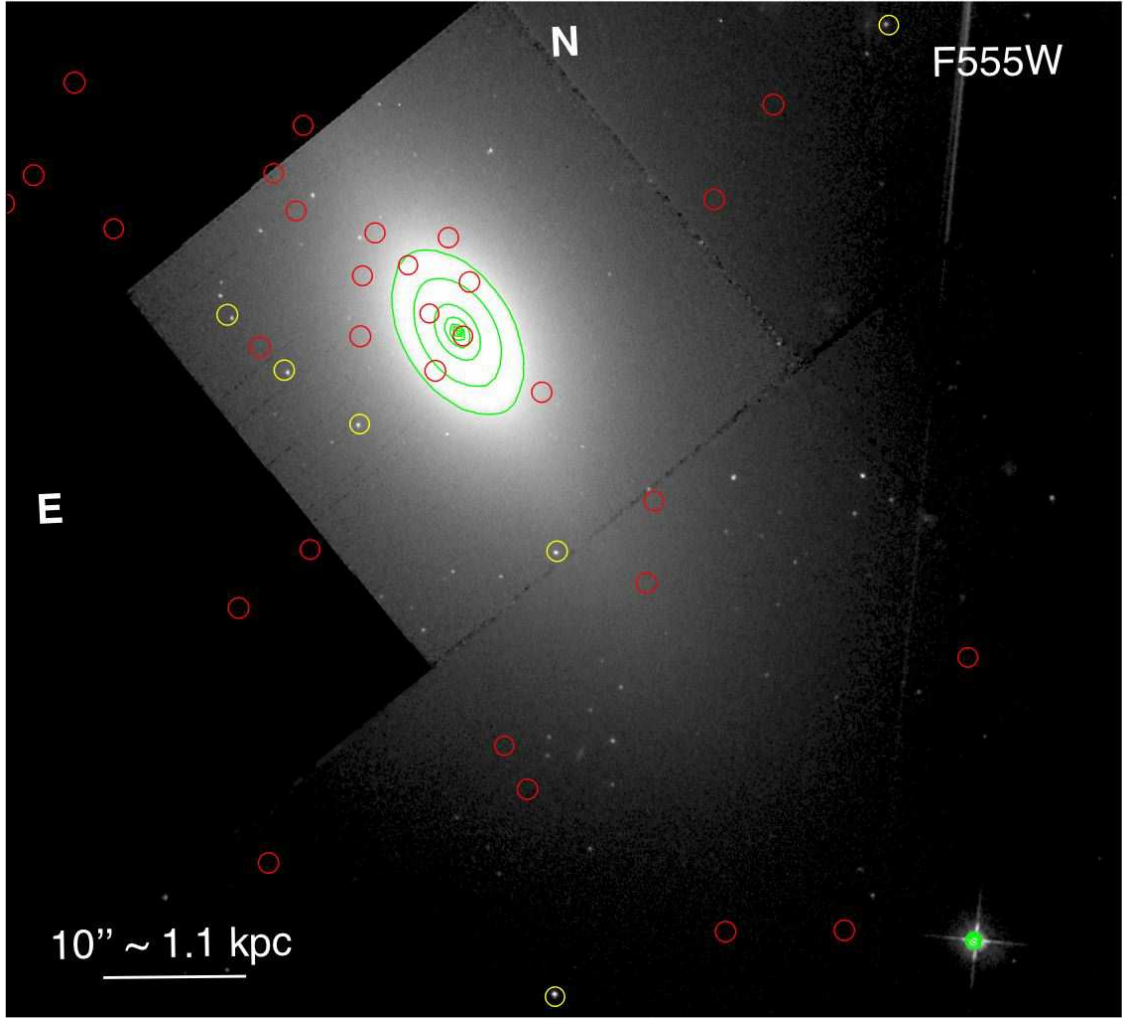}
\caption{$HST$ WFPC2 image of NGC821, with the positions of the
$Chandra$ ACIS-S sources detected by $wavdetect$ marked with red
circles.  The six yellow circles show the best optical/X-ray coincidences
falling within the D25 ellipse (not including the nuclear source) that
were found in Sect.~\ref{kimkun} and were also
used for the astrometric analysis of Sect.~\ref{nucl}.  $HST$
isophotes for the central galactic region are also overlaid.}
\label{hstimage2} 
\end{figure}

In addition, we compared the $Chandra$ and $HST$ positions by using
the archival WFPC2 F555W and F814W filter images (also used by Kundu \&
Whitmore 2001, Sect.~\ref{kimkun}). First we improved the astrometry
of the WFPC2 images, re-fitting the coordinate grid to all sources
with positional errors $\la 0\farcs3$ in the 2MASS All-Sky Catalog of
Point Sources (Cutri et al. 2003) and the USNO-B1.0 Catalog (Monet et
al. 2003).  We found 5 coincidences between the WFPC2 and the
USNO-B1.0 sources, two of which are also 2MASS sources; with these, we
improved the $HST$ absolute astrometry to $\la 0\farcs2$ in the
nuclear region.  We then compared the recalibrated WFPC2 images with
the $Chandra$ image, and identified eight sources with optical/X-ray
coincidences, not including the nuclear source S2
(Fig.~\ref{hstimage2}). Of these, seven lie inside the D25 ellipse and
have been already picked up in Sect.~\ref{kimkun} as associated with
globular clusters, with an offset smaller than the uncertainty in the
X-ray position (Tab.~\ref{kim}); however, for the present astrometric
analysis, we excluded one of them (CXOU J020817.7$+$105907) located
almost at a chip edge. The remaining six coincidences are shown with
yellow circles in Figure~\ref{hstimage2}.  They have nearly pointlike
appearance, as expected for GCs at the distance of NGC821,
and their V-magnitudes range from 20.9 to 22.4, i.e.,
they are among the brightest GCs of NGC821 (Kundu \& Whitmore 2001, as
also discussed in Sect.~\ref{kimkun}).  The eighth source with
optical/X-ray coincidence is much brighter (by almost one magnitude
in the V-band) than the brightest GC,
and lies outside the D25 ellipse (it is located at
RA=$02^h$ $08^m$ $17^s $\hskip-0.1truecm.72, Dec=$+11^{\circ}$
$00^\prime$ $31\farcs22$); it is unresolved and could be a background
AGN, as there is a higher chance of a background source outside the
optical image.

\begin{figure}[ht]
\epsscale{.9}
\plotone{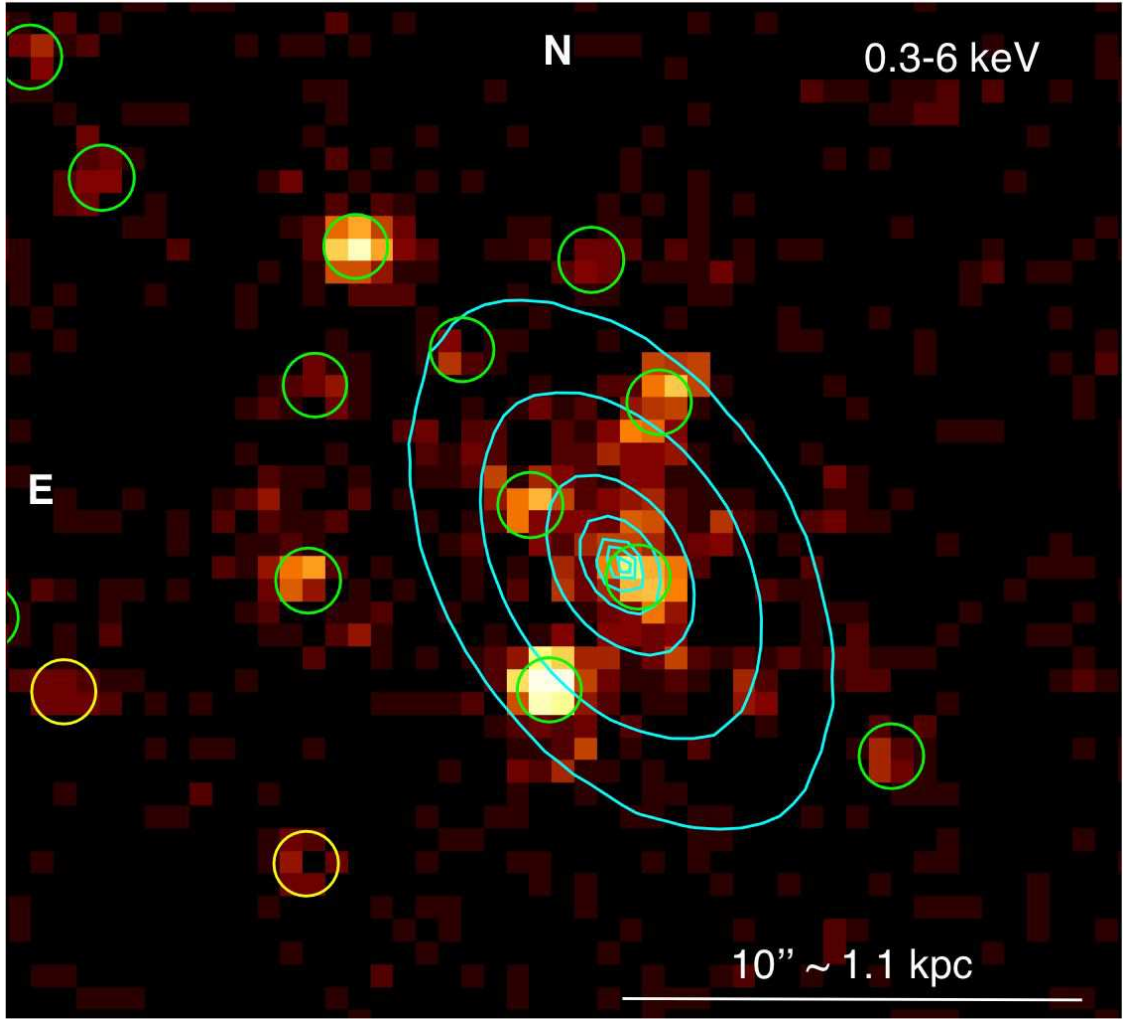}
\caption{ACIS-S image of NGC821 with green circles of $0\farcs7$ radii
marking the positions of the CIAO $wavdetect$ sources (1px=$0\farcs5$), 
and the yellow circles corresponding to GCs (as in
Fig.~\ref{hstimage2}). The galactic isophotes 
are also overlaid. In the
recalibrated WFPC2 image, the  galactic center is located at
RA=$02^h$ $08^m$
$21^s$\hskip-0.1truecm.13, Dec=$+10^{\circ}$ $59^\prime$ $41\farcs8$
(see Sect.~\ref{nucl}) and falls within the
circle of the $Chandra$ source S2.}\label{hstimage1} \end{figure}

There is no systematic shift or rotation between the revised $HST$ and
$Chandra$ positions. For each of the seven coincidences, the
displacement between the optical and X-ray positions is $\leq 0\farcs22$,
and the root-mean-square displacement for the whole sample is $\sigma
= 0\farcs14$. The relative uncertainty between $HST$ and $Chandra$ is
therefore good to $0\farcs14$, which suggests that the Chandra
pointing was in fact very accurate.  In the recalibrated WFPC2 image
the optical nucleus is located at RA=$02^h$ $08^m$
$21^s$\hskip-0.1truecm.13, Dec=$+10^{\circ}$ $59^\prime$ $41\farcs8$,
that is within $0\farcs2$ of the 2MASS position. As shown by
Fig.~\ref{hstimage1}, the WFPC2 nucleus is $0\farcs5$ off the
$wavdetect$ center of S2, a displacement larger than the average
$0\farcs14$ offset, which can be attributed to the fact that S2 is 
an extended source. In conclusion,
within the accuracy of the best relative astrometry available, S2 is
coincident with the optical center.

\subsection{Spectral analysis of sources S1--S4}\label{specnucl}

We extracted the spectra of sources S1--S4 from the merged event file,
using the extraction regions shown in Fig.~\ref{s1-s4}; the background
spectrum was derived from a source-free region outside the optical
body of the galaxy but still inside the ACIS-S3 chip. We also
generated instrumental response files, weighted by the exposure time
of each individual observation. Prior to fitting, we binned the
spectra in order to have at least 20 total counts per energy bin, with
the exception of the weakest source (S4) that was rebinned at 10
counts per energy bin\footnote{When grouping with different criteria,
or using the Cash statistics instead of the $\chi^2$ one, the spectral
results for this source do not change significantly.}.  We then fitted
the spectra with an absorbed power-law model [XSPEC model: {\it
wabs(powerlaw)}] and an optically thin thermal model corrected for
absorption [XSPEC model: {\it wabs(apec)}, with abundance Z=0.5 in
solar units, and the solar abundance values of Anders \& Grevesse
1989]; the absorption was let to be a free parameter.  The low number
of counts did not allow for fits with composite spectral models.  The
results are summarized in Table~\ref{tabspec}.

\begin{figure*}[ht]
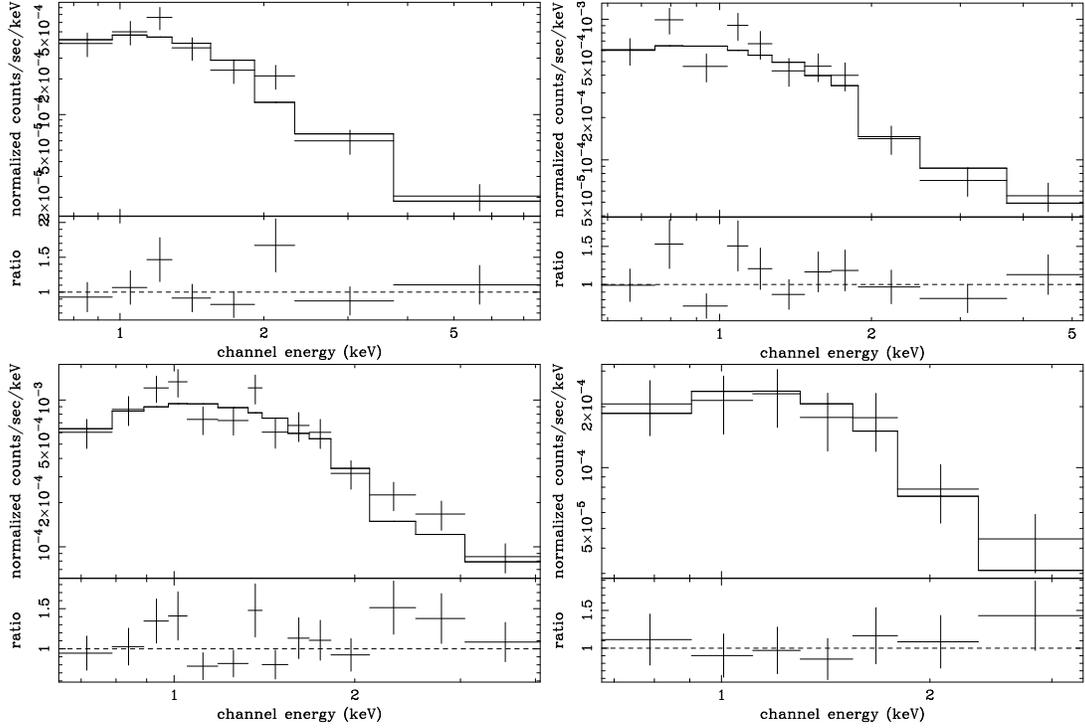

\centering
\includegraphics[scale=.3,angle=-90]{f10a.eps}
\includegraphics[scale=.3,angle=-90]{f10b.eps}\\
\includegraphics[scale=.3,angle=-90]{f10c.eps}
\includegraphics[scale=.3,angle=-90]{f10d.eps}
\plotfiddle{f10d.eps}{3.0in}{270.}{0.35}{0.35}{-500}{-500}
\caption{The $Chandra$ ACIS-S spectra of the extended sources S1--S4 
together with their best fit power law models (Tab.~\ref{tabspec}).
From left to right the panels
refer to S1 and S2 (top), S3 and S4 (bottom). See 
Sect.~\ref{specnucl} for more details.}\label{extsp} 
\end{figure*}

The spectral analysis indicates a hard spectral shape (thermal
emission of $kT\ga 3$ keV or power law emission with $\Gamma
=1.4-2.3$), with little or no intrinsic absorption.  The resulting
column densities $N_H$ are consistent with the Galactic value
($6.2\times 10^{20}$ cm$^{-2}$) for most of the fits in
Table~\ref{tabspec}; if $N_H$ is fixed to be Galactic, the best
fit spectral parameters are substantially unchanged within the
uncertainties. The spectra of S1--S4 along with their best fit power law models
are shown in Fig.~\ref{extsp}. The spectral properties of S1--S4 are overall
consistent with those typical of LMXBs (see, e.g.,
Sect.~\ref{kimkun}), even though these sources look extended
(Sect.~\ref{match}).  The source at the nucleus (S2) seems to be
harder than the other sources.

Table~\ref{tabspec} lists the source luminosities calculated from the
best fit parameters. S2 has a luminosity of about 6$\times 10^{38}\rm
erg~s^{-1}$; since it is extended, a nuclear point source, if
contributes to the emission of S2, will have a smaller luminosity.
Since we have now established that S2 is coincident with the nucleus
(Sect.~\ref{nucl}), it is reasonable that a fraction of the S2
emission originates from the MBH.  However, an estimate of this
fraction would not be unique, because it will require some model
assumptions (e.g., one could model S2 as a point source plus some
extended emission in various ways). F04 derived a 3~$\sigma$ upper
limit for the nuclear luminosity of $L_X < 4.6 \times 10^{38}~ \rm
erg~s^{-1}$ in the 0.3--10 keV band; they assumed that all of the
detected emission in the central S-shaped feature (Sect.~\ref{match})
was truly extended, and calculated the limit from the largest emission
found in a $1^{\prime\prime}\times 1^{\prime\prime}$ sliding cell
moved across this feature.
In order to avoid model dependent results, we adopt as
upper limit to the MBH emission the value derived with the same
procedure used in F04, with the
$1^{\prime\prime}\times 1^{\prime\prime}$ cell slid over and around
S2. This gives a $3\sigma$ upper limit of $L_X < 2.8 \times 10^{38}
\rm erg~s^{-1}$ in the 0.3--8 keV band, and $L_X < 1.8 \times 10^{38}
\rm erg~s^{-1}$ in the 2--10 keV band.

\section{The diffuse emission and the hot ISM 
contribution}\label{unres}

Estimating the amount of gaseous emission in NGC821 is important,
because this hot gas could be a source of fuel for the nuclear MBH
(see Sect.~\ref{intro}). The ACIS-S
adaptively smoothed mapped color image of the NGC~821 field
(Fig.~\ref{truecol}) shows diffuse emission at the center, roughly
aligned with the major axis of the galaxy. However, this diffuse
emission is not by itself proof of the presence of a hot ISM, since an
important contaminant in a galaxy like NGC821, that has a low hot gas content
(Sect.~\ref{intro}), is the undetected
portion of the LMXB population (see for example the case of NGC1316,
Kim \& Fabbiano 2003); another contaminant could be the emission of
normal stars (first discussed for this type of galaxies in
Pellegrini \& Fabbiano 1994).

In this Section we report the results of the analysis of this diffuse
emission. First, we analyze the merged data to estimate its spatial
and spectral properties (Sections~\ref{radpr} and \ref{specdiff}).
Since our deep data allow for an accurate characterization of the LMXB
population (Section~\ref{binpop}), we then use these results to
estimate how much of the diffuse emission could come from unresolved
LMXBs below the detection threshold, comparing: (1) the
shape of the radial profiles observed for the diffuse emission, the
resolved point sources and the stellar optical light
(Sect.~\ref{profiles}); (2) the radial profile of the soft emission
contributed by LMXBs and the total observed soft profile
(Sect.~\ref{soft_exp}); (3) the total luminosity of the diffuse
emission with that expected from unresolved LMXBs, using the XLF
derived in Sect.~\ref{xlf} (Sect.~\ref{xlfunres}).  Finally, in
Sect.~\ref{stars}, we derive the expected contribution from other types of
stellar sources.

\subsection{Radial profile}\label{radpr}

\begin{figure}[ht]
\epsscale{.8}
\plotone{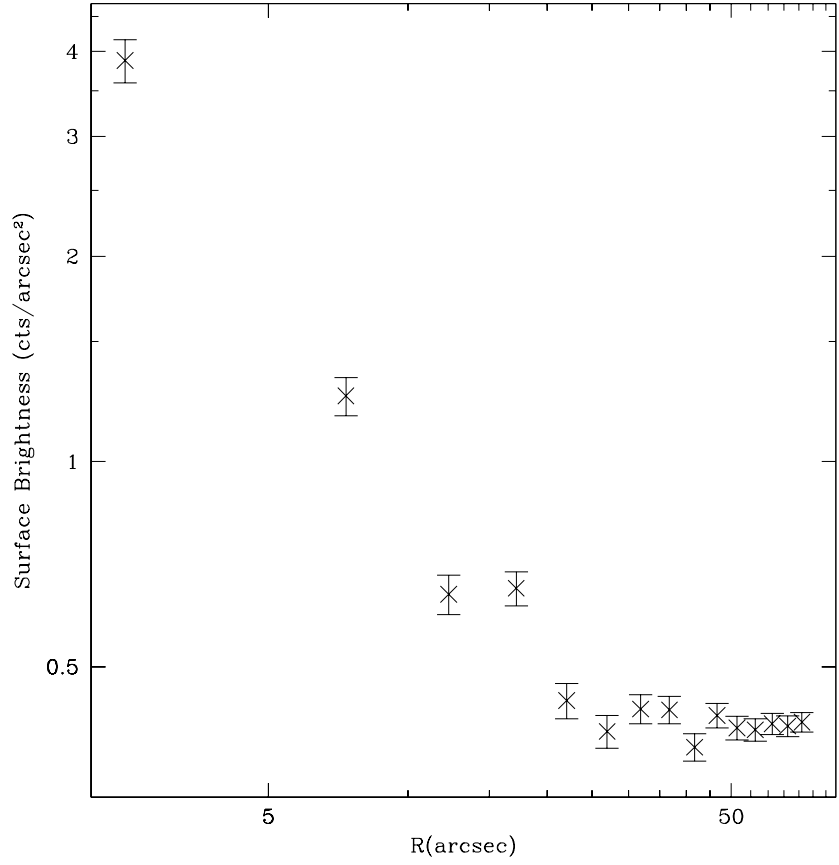}
\caption{Radial profile of the (0.3--6 keV) 
emission, after removal of sources detected by $wavdetect$ (Sect.~\ref{radpr}).
The flattening at 
radii $\ga 25^{\prime\prime}$ is due to field background. Vertical bars
give the $\pm 1\sigma$ uncertainty.}\label{rprofile}
\vskip 0.5truecm
\end{figure}

After having excluded all the detected sources, we derived a radial
profile of all the remaining counts, in the 0.3--6 keV energy band; at
energies $>6$ keV the background contribution becomes dominant and the
addition of these photons would only increase the errorbars. The
profile was centered on the nucleus, and included 30 circular annuli
(each $\sim 5^{\prime\prime}$ wide).  The resulting radial profile is shown in
Figure~\ref{rprofile}, where the presence of emission is evident as
far as $\sim 20^{\prime\prime}-30^{\prime\prime}$ from the nucleus
($\sim 2.3-3.5$~kpc at the galaxy distance). The
flattening at larger radii is consistent with the expected background
level.  Therefore, diffuse emission is detected out to a radius that
lies well within the optical extent of the galaxy, even within one
effective radius (Tab.~\ref{mainlog}).

\subsection{Spectral analysis of the diffuse emission
\label{specdiff}}

We analyzed the spectra of three circles centered on the galactic
center, of progressively larger radius: $5^{\prime\prime}$ and
$10^{\prime\prime}$, to investigate the gas presence in the
circumnuclear region, and $30^{\prime\prime}$, that is the outermost
radius at which diffuse emission is detected. Prior to their
extraction, we removed regions including detected sources.  The
background was estimated from a source-free circular area of
$50^{\prime\prime}$ radius, located outside the galaxy.  We fitted the
spectra with single component models, such as a power law (XSPEC model
$wabs * pow$) as expected from the integrated LMXB emission, and a
thermal APEC model (XSPEC model {\it wabs*apec}), to account for the
emission of a hot gas component; we then used a composite power-law
and APEC model ({\it wabs(apec+pow)}).  The results are summarized in
Tab.~\ref{denstemp}; errors quoted below give the 68\% confidence
interval for one interesting parameter.

The spectrum of a central circle of $5^{\prime\prime}$ radius is well
fitted by a single absorbed power law model, with a photon index
$\Gamma =1.77\pm 0.17$ and $N_H$ consistent with the Galactic value,
which is in good agreement with the spectral parameters of LMXBs
(Sect.~\ref{kimkun}).  The thermal model also gives an acceptable fit,
provided that the temperature is high ($kT\sim 5$ keV at the best
fit), which again is what is expected for LMXBs (Fabbiano 2006).  Next
we attempted the {\it wabs(apec+pow)} spectral model, to probe for the
presence of a hot gas component, in addition to the LMXB emission that
is clearly dominant.  $N_H$ was fixed at the Galactic value and the
abundance to the solar value, that is an average for the stellar
population in the central galactic region (Proctor et al. 2005).  At
the best fit, $\Gamma=1.61^{+0.23}_ {-0.21}$ and
$kT=0.21^{+0.16}_{-0.20}$ keV; however, this fit does not represent a
statistically significant improvement with respect to the simple power
law model, as was established via a "calibration" of the F-test
through simulations, following the prescriptions\footnote{The basic steps of the procedure consist of fitting
the observed data with the simple (pow) and composite (apec+pow)
models, getting the F-statistic for these fits, simulating a large
number spectra for the simple best fit pow model with XSPEC, fitting
each of the simulated data sets with both the simple and the composite
model and recording the F-statistic, getting the distribution of the
simulated F-statistic to compute a $p$-value (Protassov et al. 2002)
that establishes whether the addition of the thermal spectral
component is significant or not.} discussed in Protassov et al. (2002; their
Sect. 5.2).  The analysis of the spectrum of a central circle of
10$^{\prime\prime}$ radius gave similar results (Tab.~\ref{denstemp}).
This spectrum is well described by a power law consistent with the
average spectrum of LMXBs ($\Gamma=1.81\pm 0.13$ and Galactic $N_H$),
and again a thermal component is not statistically required.

We then analyzed the spectrum of the diffuse emission from the whole
galaxy (i.e., from within a radius
of $30^{\prime\prime}$).  Again, a simple power law model gives an
acceptable fit, with $\Gamma=1.74\pm 0.14$. If a thermal component is
added, with solar abundance and Galactic $N_H$, then $\Gamma
=1.59^{+0.09}_{-0.17}$ and $kT=0.59\pm 0.18$ keV; however, this thermal
component is not required statistically. The 68\% confidence upper
limit on its luminosity is L(0.3--8 keV)$<1.34\times 10^{38}$ erg
s$^{-1}$.  Based on the shallower pointings of 2002, David et
al. (2006) estimated a 90\% confidence upper limit on L(0.5--2 keV)
of $7.0\times 10^{38}$ erg s$^{-1}$, for a thermal component.

In the analysis above the presence of a small amount of thermal
emission confined to the central region might be overwhelmed by the
emission of unresolved LMXBs along the line of sight. Therefore we
also attempted the deprojection of the spectral data by using the
technique implemented within XSPEC, where the spectra extracted from
annuli centered on the nucleus are compared with the spectra expected
from the superposition along the line of sight of the emission coming
from the corresponding spherical shells [as successfully done for the
Sombrero bulge, Pellegrini et al. (2003b)].  The central region was
therefore divided in an inner circle of $5^{\prime\prime}$ radius, a
surrounding annulus with outer radius of 10$^{\prime\prime}$, and
another surrounding annulus of outer radius of
$20^{\prime\prime}$. The spectra of each of these regions were jointly
fitted with the spectral models used above. Unfortunately, the
outermost annulus did not have enough counts to add meaningful
spectral information, therefore only the two inner regions were
deprojected.  The simple power law model ({\it projct*wabs*powerlaw}),
with the photon index constrained to be the same in the two regions,
gives an acceptable fit for $\Gamma=1.68^{+0.26}_{-0.15}$ and
$N_H=2.5^{+8.0}_{-2.5}\times 10^{20}$ cm$^{-2}$ (inner region) and
$<4.3\times 10^{20}$ cm$^{-2}$ (outer region). The addition of a
thermal component [XSPEC model {\it projct*wabs(apec+pow)}, with solar
abundance and Galactic $N_H$] gives a best fit
$\Gamma=1.73^{+0.17}_{-0.15}$, and temperatures of
$kT=0.17^{+0.06}_{-0.07}$ keV for the outer annulus, and
$kT=0.03^{+0.09}_{-0.03}$ keV for the inner circle. The latter value
is too small to be meaningful, therefore the fit was repeated with a
fixed $\Gamma =1.7$ (typical of LMXBs, Sect.~\ref{kimkun}).  The outer
temperature remained unchanged and a more meaningful temperature was
obtained for the inner circle ($kT=0.18$ keV), but it was basically
unconstrained. Again, the addition of the thermal component was not
statistically significant.  At the best fit, the luminosity of the
thermal component is $\sim 5$\% and $\sim 13$\% that of the power law,
respectively for the inner and outer regions.

In summary, the analysis of the projected regions and of the central
deprojected regions never requires the presence of a soft thermal
component at a statistically significant level. The luminosity of a
soft thermal component inserted in the fit is much lower than that of
the power law ($< 1/10$ at the best fit) for the spectra of the
$5^{\prime\prime}$, $10^{\prime\prime}$ and $30^{\prime\prime}$
circles.

\subsection{Radial profiles of detected sources, diffuse emission
and galactic optical emission}\label{profiles}

\begin{figure}[ht]
\epsscale{1.}
\hskip -1.1truecm
\plotone{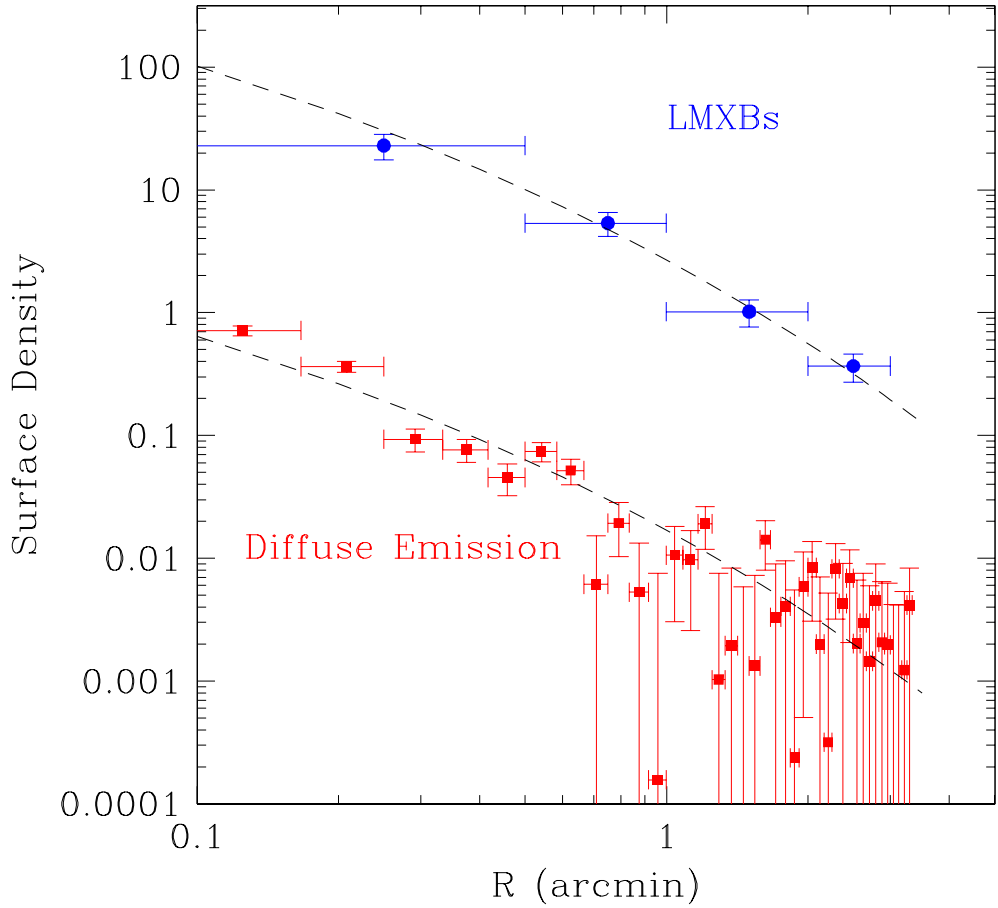}
\vskip-3truecm
\caption{Radial profiles of: detected point sources (in blue), in units of
number/square arcmin; background-subtracted
diffuse emission over 0.3--6 keV (in red), in units of counts/pixel;
galactic R-band emission (dashed lines, arbitrarily normalized). See 
Sect.~\ref{profiles} for details.}\label{Fig4}
\end{figure}

Figure~\ref{Fig4} shows the comparison of the radial profiles of three
quantities:
 (1) the background-subtracted diffuse emission in the 0.3--6 keV
 band, (2) the number density of resolved point sources, (3) the R-band
emission of NGC821. The latter, that gives the distribution of
 the stellar surface brightness, is well described by a de Vaucouleurs
(1948) law $I(R) \propto exp \left\{ -7.669*[(R/43\farcs9)^{1/4} -1]
 \right\}$, where $R$ is in arcseconds (Soria et al. 2006b).

All three profiles follow the same radial trend.  The agreement
between the radial profile of resolved point sources and that of the
stellar light is not surprising (see, e.g., Kim \& Fabbiano 2003,
Fabbiano 2006).  The similarity of these profiles with that of the
diffuse emission provides instead support to (or at least is
consistent with) the idea that also the diffuse emission mostly comes
from undetected LMXBs.

\subsection{Simulated (LMXB) and observed soft radial profiles}\label{soft_exp}

If the diffuse emission is mostly due to unresolved LMXBs, as
suggested by the analysis of the previous two Sections~\ref{specdiff}
and ~\ref{profiles}, we would expect its spectrum to have a power-law
shape typical of the LMXB population, at all radii.  Any deviation is
suggestive of a localized additional hot ISM component.

\begin{figure}[ht]
\epsscale{.9}
\plotone{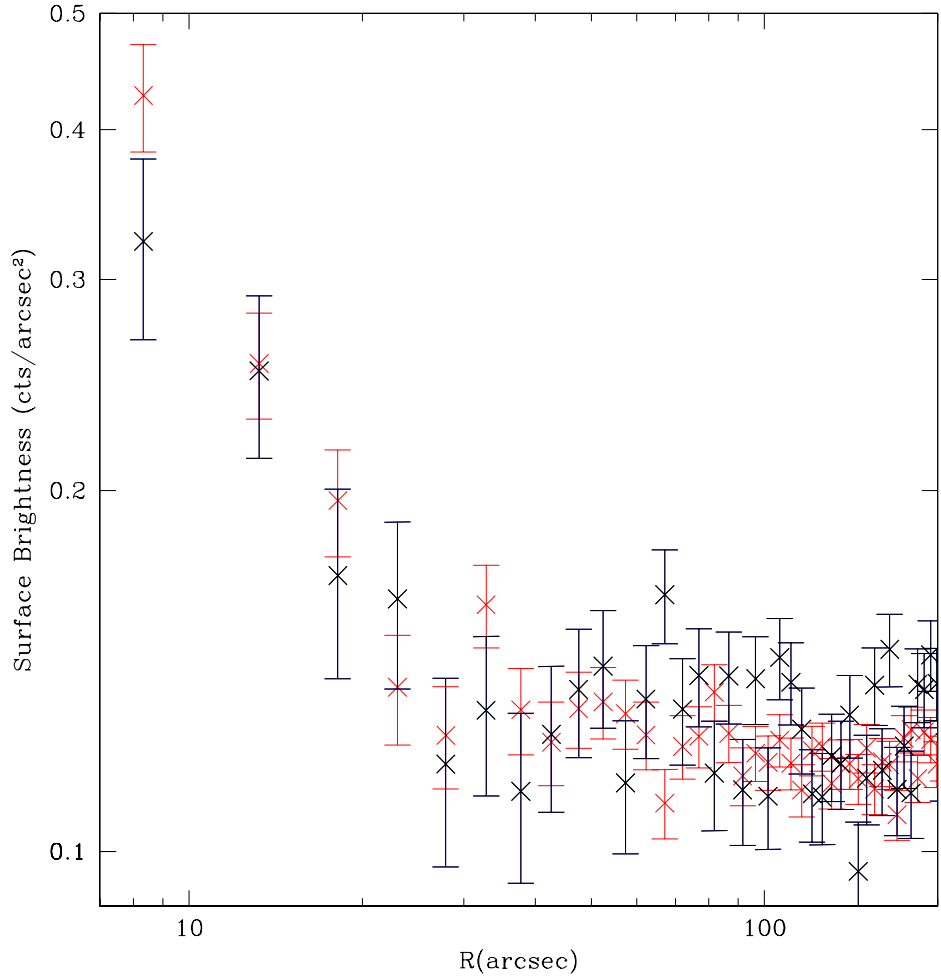}
\caption{Radial distribution of the diffuse 0.3--1.5 keV emission (in black)
that is implied
by the observed 1.5--6 keV emission, when assuming for it the spectral model
described in Sect.~\ref{soft_exp}. In red 
the observed profile in the same 0.3--1.5 keV band is also shown.
The flattening at large radii is due to field background. Vertical bars
give the $\pm 1\sigma$ uncertainty.
}\label{exp_soft}
\end{figure}

In order to further check the lack of a soft thermal component
indicated by the previous analysis at all radii, we made the following
test. We derived from observations the radial profile of the diffuse
hard emission in the 1.5--6 keV band; 
in the hypothesis that it is produced by a distributed
source whose spectrum is a power law of $\Gamma=1.7$ and Galactic
$N_H$, we then used it
to derive a "simulated" radial profile in the 0.3--1.5 keV
band.  The chosen spectrum represents the emission of unresolved
LMXBs, in accordance with the results of Sects.~\ref{kimkun}.  In
Figure~\ref{exp_soft} we compare the simulated soft profile with the
observed profile of the diffuse emission in the same energy band.  The
agreement between the two profiles is good, consistent with the idea
that the observed soft emission is due substantially to unresolved
binaries. Only within the central circle of $10^{\prime\prime}$ radius
an additional contribution from a soft source could be present, but is
not revealed by the spectral analysis (Sect.~\ref{specdiff}).

\subsection{The unresolved emission implied by the XLF}\label{xlfunres}

The emission from undetected LMXBs can be recovered from the XLF
derived in Sect.~\ref{xlf} specifically for NGC821, as the difference
between the total luminosity expected from LMXBs based on the XLF,
$L_X$(total), and the luminosity due to all detected sources,
$L_X$(detected).  In the 0.3--8 keV band, $L_X$(detected)$=6.5\times
10^{39}$ erg s$^{-1}$ for the sources falling within the D25 ellipse. For
the same region, $L_X$(total) is not just given by an integration of
the XLF, since the XLF was derived for the D25
ellipse with a central circle of $10^{\prime\prime}$ radius excluded
(Sect.~\ref{xlf}).
$L_X$(total) for the whole D25 ellipse is then obtained rescaling the
luminosity resulting\footnote{This integration assumes that the
differential XLF has the same slope ($\sim 2$) down to $L_X=10^{37}$
erg s$^{-1}$, and is done over the range $L_X=(10^{37}-10^{39})$ erg
s$^{-1}$.  Given the slope value, sources of luminosities below
$10^{37}$ erg s$^{-1}$ do not contribute significantly to the result;
for example, decreasing the lower boundary by a factor of two would
increase the resulting total luminosity by $\sim 15$\%. In addition, a
low luminosity break in the XLF is also expected (e.g., Kim et
al. 2006).} from the XLF by
the ratio between the luminosity of detected sources in the two
regions (i.e., the D25 ellipse and this ellipse without the central
circle). This procedure relies on the assumptions that (1) the XLF for
$R<10^{\prime\prime}$ and $R>10^{\prime\prime}$ is the same, and (2)
$L_X$(total)/$L_X$(detected) is also the same in these two regions.
While (1) is reasonable, (2) is less valid, since more sources would
be hidden in the inner region.  Therefore, both $L_X$(total) and
$L_X$(undetected)=$L_X$(total) -- $L_X$(detected) are actually lower
limits. For the D25 ellipse, $L_X$(total) is $7.6\times 10^{39}$ erg
s$^{-1}$, and $L_X$(undetected)$=1.1\times 10^{39}$ erg s$^{-1}$.

It is interesting now to compare the unresolved LMXBs emission with
the luminosity of the diffuse emission derived from the spectrum of a
central circle of $30^{\prime\prime}$ radius (i.e., 1.4$\times
10^{39}$ erg s$^{-1}$, Sect.~\ref{specdiff}). Therefore we repeated
the procedure above to calculate $L_X$(undetected) for this circle,
and it turned out that 
$L_X$(undetected)=$7.4\times 10^{38}$ erg s$^{-1}$. This
value is slightly lower than the spectral luminosity of the
diffuse emission; however, as noted above, it is likely an
underestimate of the true value. Also, considering the uncertainty in
the normalization of the XLF ($k=8.4^{+3.1}_{-2.5}$, Sect.~\ref{xlf}),
$L_X$(undetected) can be as large as $2.7\times 10^{39}$ erg s$^{-1}$.

In summary, for the same circle of $R=30^{\prime\prime}$, the
spectral luminosity of the diffuse emission and the XLF-based
unresolved luminosity are close, and they agree within the
uncertainties.

\subsection{Other stellar sources}\label{stars}

Since we are constraining the origin of the diffuse emission in NGC821
better than it has been possible so far in previous analyses of
low $L_X/L_B$ ellipticals, we check also for the amount of diffuse
emission that can be accounted for by stellar sources other than
LMXBs. In the old stellar population of an elliptical galaxy these
include coronae of late type main sequence stars, RS Canum Venaticorum (RS CVn)
systems and supersoft X-ray sources (Pellegrini \& Fabbiano 1994). The
latter are characterized by black body emission with effective
temperatures of 15--80 eV, X-ray luminosities up to few$\times
10^{38}$ erg s$^{-1}$ and are more frequent in late type and irregular
galaxies (Kahabka \& van den Heuvel 1997; Di Stefano \& Kong
2004). NGC4697 is the only elliptical where they
 have been identified so far (Sarazin et al. 2001). In the X-ray
 color-color plot of NGC821 (Fig.~\ref{colors}), supersoft sources
 would fall in the upper right corner, and there is no such source
 with $>30$ counts; one or two sources with less than 30 counts have a high
 C32 color, but with a large error. The contribution of supersoft
 sources will not be considered further here.

Stellar coronae of main sequence stars are all sources of thermal
X-ray emission ($L_X\approx 10^{26}-10^{30}$ erg s$^{-1}$) indicative
of coronal plasmas at temperatures of $\sim 10^6-10^7$ K (e.g.,
Schmitt et al. 1990).  These luminosities are faint in comparison to
accretion powered stellar sources, but main sequence stars of spectral
types G, K and M are present in a very large number (hereafter
respectively $N_G$, $N_K$ and $N_M$).  These $N_i$ can be recovered
from the initial stellar mass function (IMF) $\psi(M)=AM^{-2.35}$
(Salpeter 1955), where the scale-factor $A=1.67L_B$, with $L_B$ in
$L_{B,\odot}$, for a 12 Gyrs old stellar population with a 0.5 solar
metallicity (Maraston 2005), as suitable for NGC821 (Proctor et al.
2005). Then $N_i = A \int _{M_{inf}}^{M_{sup}} M^{-2.35}dM$, where
$M_{inf}$ and $M_{sup}$ are respectively 0.7$M_{\odot}$ and the main
sequence turn-off mass of 0.9$M_{\odot}$ (as appropriate for NGC821,
Maraston 2005) for G stars; 0.5 and 0.7$M_{\odot}$ for K stars; and
0.1 and 0.5$M_{\odot}$ for M stars. Integration gives $N_G=0.6 L_B$,
$N_K=1.1 L_B$ and $N_M=24.5 L_B$, therefore the collective M dwarf
X-ray emission is by far the most important, because the average X-ray
luminosity of a G, K and M star (hereafter $L_{X,M}$) are comparable
(Kuntz \& Snowden 2001). Taking $L_{X,M}\la 1.4\times 10^{27}$ erg
s$^{-1}$ in the 0.3--8 keV band\footnote{The $\la$ sign comes from
assuming thermal emission of $kT=1$ keV when converting $L_{X,M}$ in
the $ROSAT$ band (Kuntz \& Snowden 2001) to the 0.3--8 keV band. The
proper spectral description consists of two thermal components: one
with $T\sim 2-4\times 10^6$ K and the other with $T\sim 10^7$ K; the
latter usually has larger (and often much larger) emission measure
(Giampapa et al. 1996).}, the collective M dwarfs emission in NGC821
is then $N_M\times L_{X,M}\la 6.4\times 10^{38}$ erg s$^{-1}$. From a
central circle of $30^{\prime\prime}$ radius, for the de Vaucouleurs
optical profile of Sect.~\ref{profiles}, the M dwarfs emission is $\la
2.5\times 10^{38}$ erg s$^{-1}$,  that is $\la 18$\% of the luminosity 
derived from the spectrum of the diffuse emission from the same region
(Sect.~\ref{specdiff}). However, recent claims favor an IMF flatter
than the Salpeter one at the low mass end, as for the Kroupa (2001)
IMF, which is $\propto M^{-1.3}$ for $0.08\leq M/M_{\odot}\leq
0.5$. In this case the scale-factor $A=1.69 L_B$ (Maraston 2005), the 
integration gives $N_M=4.3L_B$, and the collective M dwarfs emission 
becomes $\la 1.1\times 10^{38}$ erg s$^{-1}$ for the whole galaxy, and 
$\la 0.5\times 10^{38}$ erg s$^{-1}$ for a circle of $30^{\prime\prime}$ 
radius. Therefore, it can account for $\la 3$\% of the luminosity of the 
diffuse emission.

RS CVn systems are chromospherically active objects, consisting of a G
or K giant or subgiant, with a late type main sequence or subgiant
companion (Linsky 1984). They are the most X-ray luminous late type
stars ($L_X\sim 10^{29}-10^{31.5}$ erg s$^{-1}$), and their spectra
can be well modeled by a thermal plasma with two temperatures (of
average $kT=1.3$ keV and $0.18$ keV, Dempsey et al. 1997). From
Dempsey et al.'s Fig. 3, the average 0.3--8 keV luminosity of a system
is\footnote{Again the $\la$ sign comes from assuming that most of the
emission is due to the higher temperature component.} $\la 3\times
10^{30}$ erg s$^{-1}$.  The number of RS CVn systems is a fraction of
the number of giants and subgiants (respectively $N_{giant}$ and
$N_{subg}$) expected to be present in NGC821, which
is\footnote{$N_{giant}$ and $N_{subg}$ are the product of the time
spent by stars in such evolutionary phases by the specific
evolutionary flux for a 12 Gyr old stellar population (which is fairly
independent of the IMF; Maraston 2005) and by $L_B$. Low mass metal
rich stars spend $7\times 10^8$ yrs in the G/K-subgiant phase, and
$5\times 10^8$ Gyr on the red giant branch (e.g., Renzini 1989).}
$N_{subg}+N_{giant} \approx 0.04 L_B$.  Taking half this number for
those giants or subgiants that are also in binary systems, and another
factor ($\sim 0.2$) for those that become RS CVn, then the collective
contribution is $\la 0.1(N_{subg}+N_{giant}) \times 3\times 10^{30}$
erg s$^{-1} =1.2\times 10^{28}L_B$ erg s$^{-1}$. This is $\la
2.2\times 10^{38}$ erg s$^{-1}$ for the whole galaxy, and $\la
8.9\times 10^{37}$ erg s$^{-1}$ from a circle of $30^{\prime\prime}$
radius, that is $\la 6$\% of the total diffuse emission within the
same region. A somewhat higher percentage ($\sim 15$\%) is derived in 
the 2--10 keV band when adopting the emissivity per unit stellar mass of these
sources in the Galactic plane (Sazonov et al. 2006), and assuming an average
stellar mass-to-light ratio in the B-band of $\sim 6.5$
(Sect.~\ref{hot}). This higher percentage could be in part the result of a
higher emissivity in the Milky Way, due to younger stellar ages than estimated
for NGC821 (see above).

In conclusion, the collective emission of stellar sources other than
LMXBs from the whole galaxy is $\la 3.3\times 10^{38}$ erg s$^{-1}$;
as expected, this is much smaller than the collective LMXB's emission
(7.6$\times 10^{39}$ erg s$^{-1}$, Sect.~\ref{xlfunres}), but it is
even smaller than its unresolved fraction [i.e.,
$L_X$(undetected)=$1.1\times 10^{39}$ erg s$^{-1}$,
Sect.~\ref{xlfunres}]. Also, these sources can contribute up to $\sim
10$\% of the diffuse emission detected within a radius of
$30^{\prime\prime}$ (Sect.~\ref{specdiff}).

\subsection{Conclusions on hot gas presence} 

All the different lines of investigation undertaken in this Section
consistently indicate a largely dominating LMXB contribution to the
diffuse X-ray emission.  Its spectrum does not require soft thermal
emission, that is limited to $<1.34\times 10^{38}$ erg s$^{-1}$
($<10$\% of the total diffuse luminosity) at 68\% confidence.  The
observed radial profiles of the stellar light, the resolved point
sources and the diffuse emission agree within the errors, suggesting
that the diffuse emission has a stellar origin.  The observed radial
profile in the soft band is consistent with that expected from the
sources (the LMXB population) producing the hard emission, except
perhaps for the innermost ($R\la 10^{\prime\prime}$) region. Finally,
the luminosity of undetected sources derived from the bias-corrected
XLF agrees with the spectral luminosity of the diffuse emission,
within the uncertainties. A fraction of the diffuse emission ($\la
10$\%) can also come from stellar sources other than LMXBs.

\section{Discussion}\label{disc}

In the previous Sections we have reported the results of the analysis
of a deep $Chandra$ pointing (for a total exposure of nearly 230 ks)
aimed at detecting the nuclear emission and setting stringent
constraints on the presence of hot gas both on the circumnuclear and the
galactic scale.  The deep $Chandra$ image revealed a significant
portion of the LMXB population, for which the X-ray Luminosity
Function was derived. This, together with a spectral and imaging
analysis of the diffuse emission, severely constrained the presence of
a hot ISM (Sect.~\ref{unres}). At the galactic center, in addition to diffuse
emission (which we can explain as predominantly due to unresolved LMXBs), 
we find four sources of
$L_X=(2-9)\times 10^{38}$ erg s$^{-1}$. We have
been able to establish that one of these sources (S2) is coincident
with the galactic center. Three of these source, including S2, are extended,
with one of them as long as $5^{\prime\prime}$ (S1;
Fig.~\ref{s1-s4}). These sources could be truly extended, or made of one or
more point-like components, possibly also embedded in truly diffuse
emission. Their spectral shape is consistent with that of LMXBs, but
in NGC821 there are only 2 other sources as bright as these, outside
the central circle of $10^{\prime\prime}$ radius
(Fig.~\ref{Fig2}). This supports the idea that more than one LMXB is
contributing to the emission of each source.  An alternative
possibility is that they belong to a continuous complex feature, made
of LMXBs (that may also lie there due to a projection effect), a
nucleus (possibly with associated extended emission) and an
outflow/jet. S1, in particular, has a linear structure suggestive of a jet
(Fig.~\ref{s1-s4}). Our VLA observations
(Sect.~\ref{intro}) help better address these possibilities, and are
presented in the companion paper (Pellegrini et al. 2007), together
with a discussion on whether/how accretion proceeds in this
nucleus, based on all the available observational evidence.
We estimated here a $3\sigma$ upper limit to the  0.3--8 keV 
emission of a point-like source associated with the MBH
of $< 2.8 \times 10^{38} \rm erg~s^{-1}$, which makes this
one of the quietest MBH studied with $Chandra$ (e.g., Pellegrini
2005b), with $L_X/L_{Edd}<2.5\times 10^{-8}$.

In the following we discuss the possibility that there may be
undetected hot gas in NGC~821 (Sect.~\ref{hot}). Hidden hot gas on the
nuclear scale could be a source of fuel for the MBH; furthermore, we
check whether the lack of detection on the galactic scale agrees
with expectations for the evolution of the ISM in this elliptical.

\subsection{Is there hot undetected gas? }\label{hot}

The present analysis does not detect hot gas available for accretion
at the nucleus; coupled with the absence of interstellar medium
observed at other wavelenghts (Sect.~\ref{intro}), the lack of fuel
for the MBH could be the simplest explanation for the very low levels
to which the nuclear emission has been constrained.  However, an aging
stellar population continuously returns gas to the ISM, via stellar
mass losses (e.g., Ciotti et al. 1991, David et al. 1991); as a
mimimum, the circumnuclear region should be replenished with this
fuel. Below we discuss the possible presence of this ISM.

The lack of detection of hot gas on the galactic scale can be expected
from simple energetic considerations, comparing the energy made
available per unit time by type Ia supernova explosions ($L_{SN}$)
with the energy required to steadily extract from the galactic
potential well the stellar mass losses produced per unit time
($L_{grav}$; Ciotti et al. 1991), for the whole galaxy lifetime.
$L_{SN}$ comes directly from the observed SNIa's explosion rate in
ellipticals (Cappellaro et al. 1999) rescaled for $L_B$ in
Tab.~\ref{mainlog}; this rate was higher in the past (e.g., Greggio
2005).  $L_{grav}$ derives from the stellar mass loss rate and the
galactic mass profile, that can be well modelled by the superposition
of two Hernquist (1990) density profiles\footnote{This model gives a
very good approximation of the de Vaucouleurs (1948) law, that fits
the light profile of NGC821 (Sect.~\ref{profiles}). The radial
distribution of the dark haloes of ellipticals is not well constrained
by the observations.  Theoretical arguments favor a peaked profile
(Ciotti \& Pellegrini 1992; Evans \& Collett 1997). High resolution
numerical simulations (Navarro, Frenk, \& White 1996) produce a
central density distribution equal to that of the Hernquist model.},
one for the stars with total mass $M_*$, and one for the dark matter
with mass $M_h$.  This model is then tailored onto NGC821 taking the
observed $L_B$, $R_e$, central stellar velocity dispersion
(Tab.~\ref{mainlog}) and $M_h/M_*$ ratio.  Anisotropic Jeans models
reproducing the observed velocity dispersion profiles of stars and
planetary nebulae out to 5$R_e$ constrain the range of
$(M_h+M_*)/L_B$ to be 13--17 (Romanowsky et al. 2003). For a plausible
stellar mass-to-light ratio $M_*/L_B$ (Gerhard et al. 2001, Napolitano
et al. 2005) one derives $M_h/M_*\sim 1.2$.  Finally, the stellar
mass loss rate is distributed as the stars, and derives 
by summing the mass ejected by each star as a function of its mass
(Renzini \& Ciotti 1993) for a Kroupa IMF, scaled for NGC821 as
described in Sect.~\ref{stars}; this produces a total mass loss rate
of 0.25 $M_{\odot}$ yr$^{-1}$ after 12 Gyrs, in good agreement with
the values estimated for ellipticals from $ISO$ observations (Athey et
al. 2002).  The resulting $L_{SN}\sim (3-4) L_{grav}$ during the
galaxy lifetime, with $L_{SN}\sim 4 L_{grav}$ at present; therefore
type-Ia SNe have always provided the heating to drive the stellar mass
losses in a galactic wind (neglecting the effect of radiative cooling
that is not expected to be important for very low gas densities typical of
a wind).

However, this is a global energetic calculation, and the galaxy may
host regions with $L_{SN}> L_{grav}$ and regions with $L_{SN}<
L_{grav}$.  For centrally peaked mass distributions like those
described above, numerical simulations of hot gas evolution showed
that this is indeed the most frequent case (Pellegrini \& Ciotti
1998): the ISM is driven out of the outer galactic regions in a wind,
while it is inflowing within a stagnation radius that can be largely
different from galaxy to galaxy.  To explore whether this situation
applies to NGC~821, we ran hydrodynamical simulations specific for the
galaxy model described above, plus the additional gravitational
attraction contributed by its central MBH (see Pellegrini \& Ciotti
2006 for more details on the numerical code and the time-evolving
input quantities).  The central grid spacing was set to 5 pc to allow
for a better sampling of the inner regions and the gas flow evolution
was followed for $12$ Gyrs, an age comparable with the stellar age of
NGC~821 (Proctor et al. 2005). The flow kept in a partial wind all the
time, and at the end its
total $L_X\sim$few$\times 10^{36}$ erg s$^{-1}$ in the 0.3--8 keV
band, well below our observational limit on the hot gas.  
The bulk of the hot gas was outflowing, and
directed towards the center from within a radius of $\sim 25$ pc.
At the smallest radius at which the flow is well resolved (10 pc) the
mass inflow rate is $\dot M_{in}\approx $ few$\times
10^{-5}M_{\odot}$yr$^{-1}$; the gas density is $n_e\approx
10^{-3}$ cm$^{-3}$, one third of the density value calculated from the
emission measure of the "best fit" 
thermal component for the central
circle, in the deprojected analysis of the diffuse emission
(this component is not statistically required, Sect.~\ref{specdiff}). 
Note however that the inflowing region is very
small and fully included within the extent of the S2 source
(Sect.~\ref{match}).  By varying the $M_h/M_*$ ratio, the SNIa's rate,
the age of the galaxy and the stellar mass loss rate within the limits
imposed by observational uncertainties, the value of $\dot M_{in}$
keeps within the range $(2-7)\times 10^{-5}M_{\odot}$yr$^{-1}$.  The
implications for the nuclear emission coming from this estimate of
$\dot M_{in}$ are discussed in Pellegrini et al. (2007). Here we just
note that $\dot M_{in}$ ranges from $10^{-3}$ to few$\times
10^{-5}M_{\odot}$ yr$^{-1}$ in the past $\sim 10$ Gyrs, and its integration
over this time gives a total mass accreted at the center of
$\sim 2\times 10^6 M_{\odot}$, that is $2.3\times 10^{-2}$ of the
observed MBH mass (Tab.~\ref{mainlog}). Therefore, accretion of
stellar mass losses was not effective in building this MBH mass.

Finally, we comment on the fact that, by using $Chandra$ observations,
the mass accretion rate of galactic nuclei is customarily estimated
from the analytic formula of Bondi (1952), valid for spherically
symmetric accretion from a nonrotating polytropic gas with given
density and temperature at infinity (e.g., Loewenstein et al. 2001,
Soria et al. 2006a). Infinity is replaced with an accretion radius
$r_{acc}=2GM_{BH}/c_s^2$ (Frank et al. 2002), where the sound speed
$c_s\propto \sqrt{kT}$ is calculated as close as possible to the MBH.
Even in those cases where $kT$ can be observed close to the galactic
center, so that one can be confident to derive an approximate estimate
of $r_{acc}$, there are additional ingredients that must be considered
in the estimate of the mass accretion rate,
which were not included in the Bondi (1952) treatment: 1) the presence
of mass and energy sources (i.e., stellar mass losses and SNIa's
heating); 2) the presence of cooling; 3) the fact that $r_{acc}$ is
not a true "infinity" point, since the gas here experiences a pressure
gradient.  The simulations described here take into account all these
aspects, and should give a more reliable mass accretion rate than the
Bondi theory.  However, they have some limits too. At these very low
$\dot M_{in}$ values, and very small inflowing regions, the detailed shape
of the galactic mass profile becomes important. In the case of NGC821,
the observed profile is steeper than modelled here at the very center
(Gebhardt et al. 2003), which should produce a larger $\dot M_{in}$.
The true accretion rate could be somewhat higher than $\dot M_{in}$
also if even the stellar mass losses within the innermost radius
resolved by the simulations are to be accreted.  But further pursuing
the flow behavior closer to the MBH with the simulations
used here runs into a problem: the discrete
nature of the stellar distribution becomes important, since the
accretion time ($\sim 3\times 10^5$ yrs from 10 pc, in the
simulations) becomes comparable to (or lower than) the time required
for the stellar mass losses to mix with the bulk flow (Mathews 1990),
and the time elapsing between one SNIa event and the next.

\section{Summary}\label{concl}

We have observed the nearby, inactive elliptical galaxy NGC821, known
to host a central MBH, with deep $Chandra$ pointings, in order to put
strong constraints on its nuclear emission and the presence of gas
available for accretion. Our results can be summarized as follows:

\begin{enumerate}

\item We detect 41 sources within the optical size (D25) of the galaxy;
excluding 4 particular sources at the center, these represent the LMXB
population of NGC~821.  Their X-ray colors, and the
spectral analysis of the three brightest ones, are consistent with
spectral shapes described by power laws of photon index in the range
$\Gamma=1.5-2.0$, without significant intrinsic absorption.  There is
no clear evidence for supersoft sources.

\item In the overlap region between D25 and the field of view of
previously taken $HST$ WFPC2 images, there are six clear associations
of LMXBs with globular clusters, plus two additional marginal matches.
These LMXBs reside in the brightest GCs of the galaxy. Excluding the
two marginal matches, $\sim 30$\% of the X-ray sources lie in GCs, and
$\sim 6.6$\% of the GCs have a LMXB, broadly in agreement with the percentages
found for ellipticals in previous works.

\item The XLF of these sources is well fitted by a simple power law
down to $L_X=3\times 10^{37}$ erg s$^{-1}$ (or to $L_X=2\times
10^{37}$ erg s$^{-1}$ after completeness correction), with a slope
($\beta=2.0\pm 0.3$) similar to that found previously for other
ellipticals. The ratio between the total $L_X$ due to LMXBs and the
galactic $L_B$ is lower than the average found for early-type
galaxies, but still within the observed scatter.

\item At the position of the galactic center a source (S2) is detected
for the first time.  Its spectral shape is quite hard
($\Gamma=1.49^{+0.14}_{-0.13}$), without intrinsic absorption. S2 is
however slightly extended, and a 0.3--8 keV upper limit of $2.8\times 10^{38}$
erg s$^{-1}$ is derived for a point-like nuclear source, one of the smallest
established with $Chandra$.

\item Three other sources, with a spectral shape typical of LMXBs,
are detected in the central galactic region; they are  as bright
as the brightest sources in the XLF. Only one (S3) is
consistent with being pointlike; the other
two (S1 and S4) are extended, and could be due to the superposition of
few LMXBs and/or truly diffuse emission. The morphology of S1
resembles a jet-like feature.

\item Diffuse emission is detected out to a radius of $\sim
30^{\prime\prime}$. A few independent lines of investigation,
exploiting the spectral and imaging capabilities of $Chandra$,
consistently indicate that this diffuse emission is due to unresolved 
LMXBs, and
provide no evidence for hot gas, either close to the center or on a
larger scale. The spectral analysis gives 
a ($1\sigma$) limit of L(0.3--8 keV)$<1.34\times 10^{38}$ erg s$^{-1}$
on any soft thermal component within a radius of $30^{\prime\prime}$.

\item Other unresolved stellar sources (mostly M dwarfs and RS CVn
systems) could contribute to the diffuse emission by $\la 10$\%.  The
general scaling of their collective $L_X$ with the galactic $L_B$ 
was also derived.

\item Numerical simulations of the hot gas evolution for a galaxy model
tailored on NGC~821 show that the bulk of the gas is driven out in a
wind for the whole galaxy lifetime, due to the heating provided by
type Ia supernovae. While this may have been expected based on simple
energetic calculations, the simulations also show that the gas is
accreting towards the center from within a very small inner region.
This gas flow pattern is expected to be common in low
$L_B$ ellipticals, due to their cuspy central mass distribution.  
Gaseous accretion alone at the rates given by the simulations
cannot have formed the nuclear MBH of NGC821.

\end{enumerate}

In a companion paper (Pellegrini et al. 2007), we present further
observational results for the central region of NGC~821, obtained with
$Spitzer$ and the VLA, and discuss various possibilities
for the nature of the accretion process in this very low luminosity 
nucleus.

\acknowledgments

We thank F. Ferraro for useful information related to
Sect.~\ref{stars}, the referee for useful comments and A. Kundu for
kindly providing the list of globular clusters used in
Sect.~\ref{kimkun} and information on possible optical counterparts of
the X-ray sources in Tab.~\ref{kim}.  S.P. acknowledges partial
financial support from the Italian Space Agency ASI (Agenzia Spaziale
Italiana) through grant ASI-INAF I/023/05/0.  Partial support for this
work was provided by the NASA $Chandra$ Guest Observer grant GO5-6110X
and by the $Chandra$ X-ray Center NASA contract NAS8-39073. The data
analysis was supported by the CXC CIAO software and CALDB. We have
used NASA NED and ADS facilities, and have extracted archival data
from the Hubble Space Telescope archive.


\clearpage

\begin{deluxetable}{lccccccccc}
\rotate
\tablecaption{NGC~821: main properties\label{mainlog}}
\tablewidth{0pt}
\tablehead{
\colhead{Type\tablenotemark{a}} &
\colhead{B$^0_T$\tablenotemark{a}} &
\colhead{D\tablenotemark{b}}& 
\colhead{log $L_{B}$} &
\colhead{Size\tablenotemark{a}} &
\colhead{R$_{\rm e}$\tablenotemark{c}} &
\colhead{$\sigma_e$\tablenotemark{d}} &
\colhead{N$_H$\tablenotemark{e}} &
\colhead{$M_{\rm BH}$\tablenotemark{f}} &
\colhead{1$^{\prime\prime}$}     
\\
\colhead{ } &
\colhead{(mag)} &
\colhead{(Mpc)} &
\colhead{($L_{B,\odot}$)} &
\colhead{(arcmin)} &
\colhead{($^{\prime\prime}$,kpc)} &
\colhead{(km~s$^{-1}$)} &
\colhead{(cm$^{-2}$)} &
\colhead{($10^7 M_{\odot}$)}&
\colhead{(pc)}
}
\startdata
 E6 & 11.72 & 24.1 & 10.27 & 2.57x1.62 & 43.9, 5.1 & 209 & 6.2$\times 10^{20}$ & 8.5$\pm 3.5$ & 117 \\
\enddata

\tablenotetext{a}{Type, $B^0_T$ and size from de Vaucouleurs et al. (1991; 
RC3). The size gives the major and minor axis of the D25
ellipse, that is the 25.0 B-mag/square arcsec isophote. The position
angle is $25^{\circ}$ (RC3).}

\tablenotetext{b}{Distance D from Tonry et al. (2001).}

\tablenotetext{c}{Effective radius R$_{\rm e}$ in the R-band (from 
Soria et al. 2006b; see also Sect.~\ref{profiles}).}

\tablenotetext{d}{Effective stellar velocity dispersion (averaged 
over $R_e$) from Pinkney et al. (2003).}

\tablenotetext{e}{Galactic hydrogen column density
(Dickey \& Lockman 1990).}

\tablenotetext{f}{Gebhardt et al. (2003) report a value of
3.7$^{+2.4}_{-0.8}\times 10^7\,M_{\odot}$, later revised to the value
given here (Richstone et al., astro-ph/0403257) that is considered
more reliable (Gebhardt, K. 2006, private communication).}

\end{deluxetable}

\clearpage

\begin{table}[ht]
\caption{$Chandra$ ACIS-S observing log for NGC~821.\label{obslog}}
\begin{center}
\begin{tabular}{ccc}
\hline
Obs ID & Date & Exposure (ks) \\
\hline\hline
4408 & 2002, Nov 26 & 25.3 \\
4006 & 2002, Dec 01 & 13.7 \\
5692 & 2004, Dec 04 & 28.0 \\
6314 & 2005, Jun 20 & 40.1 \\
6310 & 2005, Jun 21 & 32.4 \\
6313 & 2005, Jun 22 & 50.1 \\
5691 & 2005, Jun 23 & 40.1 \\
\hline
\end{tabular}
\end{center}
\end{table}

\clearpage
\pagestyle{empty}
\begin{deluxetable}{lcccccccccccccc}
\rotate
\tablecaption{Properties of the X-ray point source population within D25
\label{kim}}
\tabletypesize{\scriptsize}
\tablewidth{0pt}
\tablehead{
\colhead{CXOU Name}&
\colhead{ RA}&
\colhead{ DEC}&
\colhead{$\Delta$}&
\colhead{R}& 
\colhead{cnt} &\colhead{err}&\colhead{ $F_X$} &\colhead{ $L_X$} &\colhead{ HR} &\colhead{ err}&\colhead{ C21} &\colhead{ err}&\colhead{C32}&\colhead{err}\\
\colhead{ (1)}   &\colhead{(2)} &\colhead{ (3)} &\colhead{(4)}  & \colhead{(5)} &\colhead{  (6)}  &\colhead{  (7)}&\colhead{  (8)} &\colhead{(9)} &\colhead{ (10)}&\colhead{(11)}& \colhead{(12)}&\colhead{(13)}&\colhead{ (14)}&\colhead{(15)} }
\startdata


   J020821.4$+$105946\tablenotemark{a} & 2 8 21.4 & $+$10 59 46.5 & 0.40 & 6.24 & 20.6 & 6.1 &  8.0 & 0.56 & $-$0.52 &($-$0.74 $-$0.27) &  0.19 &($-$0.13  0.52) &  0.29& ($-$0.09  0.75)  \\
   J020820.7$+$105937\tablenotemark{b} & 2 8 20.7 & $+$10 59 37.7 & 0.40 & 6.75 &  29.2 &  8.5 & 11.3 & 0.79 & $-$1.00 & ($-$1.00 $-$0.68) &  0.24 & (0.00  0.46) &  0.52 & ( 0.07  1.34)   \\
   J020821.2$+$105948\tablenotemark{b} & 2 8 21.2 & $+$10 59 48.4 & 0.39 & 6.95 &  21.0 &  6.1 &  8.1 & 0.56 & $-$0.52 & ($-$0.74 $-$0.27) & $-$0.43 & ($-$0.99 $-$0.04)  & 0.39 & (0.07  0.77) \\
   J020821.6$+$105941\tablenotemark{b}  & 2 8 21.6 & $+$10 59 41.5 & 0.21 & 7.16 &  77.2 & 10.5 & 29.9 & 2.08 & $-$0.53 & ($-$0.63 $-$0.43) & $-$0.19 & ($-$0.36 $-$0.11)  & 0.58 & (0.46  0.78) \\
   J020821.6$+$105945\tablenotemark{b}  & 2 8 21.6 & $+$10 59 45.7 & 0.40 & 8.16 &  20.3 &  6.1 &  7.9 & 0.55 & $-$0.36 & ($-$0.60 $-$0.10) & $-$0.51 & ($-$1.24 $-$0.05)  & 0.24 & ($-$0.07  0.59)  \\
   J020821.5$+$105948\tablenotemark{b,c}  & 2 8 21.5 & $+$10 59 48.7 & 0.15 & 9.44 &  164.9 & 14.3 & 63.9 & 4.44 & $-$0.59 & ($-$0.64 $-$0.51) & $-$0.27 & ($-$0.38 $-$0.20)  & 0.58 & (0.48  0.67)  \\
   J020821.6$+$105935\tablenotemark{a}  & 2 8 21.6 & $+$10 59 35.3 & 0.42 & 9.53 &  20.4 &  7.4 &  7.9 & 0.55 & $-$0.56 & ($-$0.90 $-$0.24) & $-$0.53 & ($-$1.44 $-$0.09)  & 0.84 & (0.38  1.85)  \\
   J020822.0$+$105939\tablenotemark{a}  & 2 8 22.0 & $+$10 59 39.1 & 0.43 & 12.74&  20.6 &  6.0 &  8.0 & 0.55 & $-$0.31 & ($-$0.54 $-$0.05) & $-$0.02 & ($-$0.37  0.30)  & 0.34 & ($-$0.01  0.76)  \\  
   J020822.1$+$105940\tablenotemark{b}  & 2 8 22.1 & $+$10 59 40.7 & 0.49 & 14.21& 15.5 &  5.3 &  6.0 & 0.42 & $-$0.64 & ($-$0.87 $-$0.36)  & $-$0.62 & ($-$1.30 $-$0.17)  & 0.74  & (0.32  1.39)  \\  
   J020821.9$+$105950\tablenotemark{b}  & 2 8 21.9 & $+$10 59 50.2 & 0.36 &  14.51 &  25.3 &  6.6 &  9.8 & 0.68 & $-$0.79 & ($-$0.96 $-$0.57) &  0.02 & ($-$0.19  0.23)  & 0.79  & (0.36  1.58)  \\  
   J020820.7$+$105926\tablenotemark{a}  & 2 8 20.7 & $+$10 59 26.6 & 0.45 &  16.39 &  23.8 &  7.3 &  9.2 & 0.64 & $-$1.00 & ($-$1.00 $-$0.94) & $-$0.46 & ($-$0.99 $-$0.12) &  1.23 & (0.76  2.10)  \\  
   J020822.2$+$105942\tablenotemark{a}  & 2 8 22.2 & $+$10 59 42.9 & 0.67 &  16.53 &  8.2  & 4.3 &  3.2 & 0.22 & $-$1.00 & ($-$1.00 $-$0.90) & $-$0.16 & ($-$0.65  0.25)  & 1.18  & (0.60  2.12)  \\  
   J020822.0$+$105952\tablenotemark{b}  & 2 8 22.0 & $+$10 59 52.8 & 0.41 &  17.38 &  19.9 &  5.9 &  7.7 & 0.54 & $-$0.58 & ($-$0.78 $-$0.34) & $-$0.22 & ($-$0.61  0.12)  & 0.41 & (0.07  0.83)  \\  
   J020820.2$+$105930\tablenotemark{b}  & 2 8 20.2 & $+$10 59 30.1 & 0.35 &  17.60 &  34.1 &  8.6 & 13.2 & 0.92 & $-$0.35 & ($-$0.51 $-$0.08) & $-$0.05 & ($-$0.36  0.23)  & 0.30  & (0.14  0.72)  \\  
   J020821.9$+$105956  & 2 8 21.9 & $+$10 59 56.2 & 0.44 &  18.42 &  42.0 & 10.5 & 16.3 & 1.13 & $-$0.53 & ($-$0.70 $-$0.38) & $-$0.43 & ($-$0.83 $-$0.14)  & 0.54 & (0.31  0.79)  \\  
   J020821.8$+$105926  & 2 8 21.8 & $+$10 59 26.5 & 0.57 &  18.43 &  10.4 &  6.2 &  4.0 & 0.28 & $-$0.14 & ($-$0.76  0.45) & $-$0.26 & ($-$1.43  0.51) & $-$0.06 & ($-$0.67  0.50)  \\  
   J020819.9$+$105951\tablenotemark{b}  & 2 8 19.9 & $+$10 59 51.3 & 0.48 &  20.05 &  11.7 &  6.1 &  4.5 & 0.31 & $-$1.00 & ($-$1.00 $-$0.64) &  0.12 & ($-$0.32  0.56)  & 0.70 & (0.06  1.93)  \\  
   J020820.2$+$105924\tablenotemark{b}  & 2 8 20.2 & $+$10 59 24.4 & 0.55 &  21.45 &  13.0 &  6.7 &  5.0 & 0.35 & $-$1.00 & ($-$1.00 $-$0.58) & $-$0.13 & ($-$0.72  0.33)  & 0.50 & ($-$0.01  1.47)  \\  
   J020822.2$+$105922\tablenotemark{c}  & 2 8 22.2 & $+$10 59 22.4 & 0.15 &  24.76 &  214.2&  16.1 & 83.0 & 5.76 & $-$0.45 & ($-$0.51 $-$0.38) & $-$0.45 & ($-$0.53 $-$0.35)  & 0.42 & (0.35  0.50) \\  
   J020822.8$+$105948  & 2 8 22.8 & $+$10 59 48.9 & 0.43 &  25.46 &  25.1 &  7.4 &  9.7 & 0.68 & $-$0.28 & ($-$0.49 $-$0.02) & $-$0.31 & ($-$0.96  0.10)  & 0.15 & ($-$0.07  0.42)  \\  
   J020819.6$+$105957\tablenotemark{b}  & 2 8 19.6 & $+$10 59 57.9 & 0.35 &  27.17 &  26.6 &  7.4 & 10.3 & 0.72 &  0.01 & ($-$0.25  0.21) & $-$0.85 & ($-$2.11 $-$0.31)  & 0.12 & (0.05  0.49)  \\  
   J020820.9$+$105912\tablenotemark{b}  & 2 8 20.9 & $+$10 59 12.9 & 0.58 &  28.80 &  13.7 &  8.1 &  5.3 & 0.37 & $-$0.36 & ($-$0.77  0.02) &  0.54 & (0.01  1.39) & $-$0.37 & ($-$1.25  0.26)  \\  
   J020823.1$+$105952\tablenotemark{c}  & 2 8 23.1 & $+$10 59 52.6 & 0.14 &  31.95 &  301.0&  18.6 &  116.7 & 8.10 & $-$0.44 & ($-$0.49 $-$0.39) & $-$0.27 & ($-$0.35 $-$0.22)  & 0.43 & (0.37  0.50)  \\
   J020820.8$+$105909\tablenotemark{b}  & 2 8 20.8 & $+$10 59  09.9 & 0.51 &  32.01 &  19.2 &  8.6 &  7.4 & 0.52 & $-$0.70 & ($-$0.86 $-$0.37) &  0.48  & (0.14  0.94)  & 0.09 & ($-$0.44  0.77)  \\  
   J020822.9$+$105959  & 2 8 22.9 & $+$10 59 59.1 & 0.27 &  32.30 &  44.6 &  8.7 & 17.3 & 1.20 & $-$0.62 & ($-$0.74 $-$0.45) & $-$0.21 & ($-$0.47  0.03)  & 0.55 & (0.37  0.92)  \\  
   J020823.3$+$105950  & 2 8 23.3 & $+$10 59 50.6 & 0.30 &  33.26 &  43.3 &  8.0 & 16.8 & 1.17 & $-$0.54 & ($-$0.63 $-$0.35) &  0.03 & ($-$0.16  0.20)  & 0.18 & (0.04  0.39) \\ 
   J020819.1$+$110003\tablenotemark{a}  & 2 8 19.1 & $+$11  00 03.6 & 0.47 &  37.02 &  11.4 &  5.9 &  4.4 & 0.31 &  0.10 & ($-$0.45  0.68) & $-$0.01 & ($-$1.03  0.87) & $-$0.35 & ($-$1.14  0.17)  \\  
   J020823.0$+$110008  & 2 8 23.0 & $+$11 00 08.9 & 0.35 &  39.19 &  30.8 &  7.7 & 11.9 & 0.83 & $-$0.31 & ($-$0.42 $-$0.02) & $-$0.95 & ($-$2.23 $-$0.39)  & 0.24 & (0.09  0.49)  \\  
   J020822.0$+$105904\tablenotemark{b}  & 2 8 22.0 & $+$10 59  04.6 & 0.44 &  39.42 &  29.0 &  7.5 & 11.2 & 0.78 & $-$1.00 & ($-$1.00 $-$0.73) & $-$0.31 & ($-$0.67 $-$0.02)  & 0.77 & (0.37  1.55)  \\  
   J020818.7$+$105919\tablenotemark{b}  & 2 8 18.7 & $+$10 59 19.3 & 0.43 &  41.70 &  26.4 &  7.4 & 10.2 & 0.71 & $-$0.56 & ($-$0.81 $-$0.29) & $-$0.02 & ($-$0.34  0.27)  & 0.52 & (0.13  1.21)  \\  
   J020819.9$+$105900\tablenotemark{b}  & 2 8 19.9 & $+$10 59  00.0 & 0.54 &  45.41 &  17.1 &  6.6 &  6.6 & 0.46 & $-$0.20 & ($-$0.61  0.21) & $-$0.57 & ($-$1.67 $-$0.06)  & 0.41 & (0.02  1.02)  \\  
   J020820.7$+$105855\tablenotemark{a}  & 2 8 20.7 & $+$10 58 55.4 & 0.30 &  46.64 &  52.8 &  9.2 & 20.4 & 1.42 & $-$0.46 & ($-$0.59 $-$0.32) & $-$0.40 & ($-$0.68 $-$0.16)  & 0.57 & (0.37  0.78)  \\  
   J020819.3$+$105900\tablenotemark{d} & 2 8 19.3 & $+$10 59  00.2 & 0.53 &  49.23 &  13.6 &  6.1 &  5.3 & 0.37 &  0.66   & (0.20  0.84) &   0.26 & ($-$0.65  1.34) & $-$0.69 & ($-$1.81 $-$0.14)  \\  
   J020824.1$+$110007  & 2 8 24.1 & $+$11 00 07.8 & 0.31 &  51.53 &  43.6 &  8.5 & 16.9 & 1.17 & $-$0.19 & ($-$0.35 $-$0.03) & $-$0.56 & ($-$1.06 $-$0.22)  & 0.26 & (0.13  0.47)  \\ 
   J020817.7$+$105907\tablenotemark{a}  & 2 8 17.7 & $+$10 59 07.7 & 0.28 &  60.28 &  60.9 &  9.6 & 23.6 & 1.64 & $-$0.72 & ($-$0.82 $-$0.60) & $-$0.36 & ($-$0.54 $-$0.20)  & 0.68 & (0.51  0.93)  \\  
   J020820.2$+$110041\tablenotemark{d} & 2 8 20.2 & $+$11 00 41.1 & 0.35 &  61.17 &  24.4 &  7.2 &  9.5 & 0.66 &  1.00   & (0.66  1.00) & $-$0.30 & ($-$1.39  0.36) & $-$0.49 & ($-$0.95 $-$0.15)  \\  
   J020822.5$+$110044  & 2 8 22.5 & $+$11 00 44.1 & 0.48 &  65.73 &  15.0 &  6.4 &  5.8 & 0.41 &  0.24   & ($-$0.29  0.72) & $-$0.73 & ($-$2.06 $-$0.09)  & 0.11 & ($-$0.27  0.57) \\  
\enddata
\tablecomments{Units of rigth ascension are hours, minutes and seconds, and units of declination are degrees, arcminutes and arcseconds. Units of $\Delta$
(the positional uncertainty) and $R$ (the distance from the galactic center)
are arcseconds. The positional uncertainty is  based on the statistical error 
of the X-ray data alone (Sect.~\ref{binpop}), while the absolute positional
error is discussed in Sect.~\ref{nucl}.
Units of $F_X$ are $10^{-16}$ erg s$^{-1}$ cm$^{-2}$ and those of $L_X$ are $10^{38}$ erg s$^{-1}$; both quantities have been absorption-corrected
for the Galactic column density and refer to the 0.3--8 keV band.
See Sect.~\ref{binpop} for more details (e.g., on the definitions of the 
hardness ratio HR and the colors C21 and C23, and the calculation of their 
errors).}

\tablenotetext{a}{The position of this source is coincident with that
of a candidate globular cluster found in the $HST$/WFPC2 images, see
Sect.~\ref{kimkun}; the offset between the optical and X-ray position
is always well within $\Delta$, except for the first of these 8
sources (J020821.4$+$105946) that has an offset of 1.24$\Delta$.}

\tablenotetext{b}{This is a field LMXB candidate (see Sect.~\ref{kimkun}).}

\tablenotetext{c}{A specific spectral analysis has been done
for this source (Sect.~\ref{kimkun} and Tab.~\ref{baldi}).}

\tablenotetext{d}{This X-ray source may correspond to a faint
background galaxy, from an inspection of the $HST$ images discussed in
Sect.~\ref{kimkun} (A. Kundu, private communication).}

\end{deluxetable}

\clearpage
\pagestyle{plaintop}
\begin{table*}[ht] 
\caption{Spectral analysis results for the brightest sources in Tab.~\ref{kim}.\label{baldi}} 
\begin{center}
\begin{tabular}{lccc} 
\hline 
           &  J020821.5$+$105948   &  J020822.2$+$105922   &  J020823.1$+$105952   \\
\hline\hline 
\tbsp
$wabs(pow)$ : & & & \\
N$_H$ (10$^{21}$ cm$^{-2}$)& $2.2_{-1.0}^{+0.8}$& $0.5_{-0.5}^{+1.3}$&  $1.5_{-0.5}^{+0.5}$\\
$\Gamma$ & $1.98_{-0.28}^{+0.42}$&  $1.21_{-0.24}^{+0.17}$& $1.58_{-0.15}^{+0.13}$ \\
$\chi^2/$dof  & 8.1/4 & 5.2/7 & 7.6/11 \\
Flux ($10^{-15}$ erg cm$^{-2}$ s$^{-1}$) &4.5 & 9.7 & 10.9 \\
Luminosity ($10^{38}$ erg s$^{-1}$) & 4.1 & 6.9 & 8.9 \\ 
\hline\hline 
\end{tabular} 

\end{center} 

\tablecomments{The $N_H$ values are in addition to the Galactic one.
Fluxes are observed values, luminosities are corrected for absorption,
and are calculated for the 0.3--8 keV band. Errors give the 68\%
confidence interval for one interesting parameter. See also Sect.~\ref{kimkun}.}

\end{table*}

\clearpage

\begin{table*}[ht] 
\caption{Spectral analysis results for the central sources S1--S4.\label{tabspec}} 
\begin{center}
\begin{tabular}{lcccc} 
\hline 
           &   S1   &   S2   &   S3   &   S4  \\
\hline\hline 
\tbsp
Net counts & 178$\pm 14$    &  246$\pm 16$   &  324$\pm 18$   &  82$\pm 9$   \\
\tbsp
$wabs(pow)$ : & & & & \\
N$_H$ (10$^{21}$ cm$^{-2}$)& $1.7_{-1.0}^{+0.8}$& $<0.5$&  $2.6_{-0.6}^{+0.5}$& $3.0_{-1.5}^{+1.3}$\\
$\Gamma$ & $1.80_{-0.23}^{+0.35}$&  $1.49_{-0.13}^{+0.14}$&  $2.23_{-0.17}^{+0.23}$ &  $2.28_{-0.34}^{+0.52}$ \\
$\chi^2/$dof  & 7.3/6 & 11.5/9 &  15.8/12& 2.1/5 \\
Flux ($10^{-15}$ erg cm$^{-2}$ s$^{-1}$) &5.8 & 8.6 & 9.1 & 2.3 \\
Luminosity ($10^{38}$ erg s$^{-1}$) & 4.7 & 6.0 & 9.0 & 1.9 \\ 
\tbsp
$wabs(apec)$ : & & & & \\
N$_H$ (10$^{21}$ cm$^{-2}$) & $<1.1$ & $<0.5$ &$0.9_{-0.5}^{+0.6}$&  $1.4_{-1.3}^{+1.5}$ \\   
$kT$ (keV) & $5.3_{-1.8}^{+4.7}$&$8.5_{-3.4}^{+11}$&$4.3_{-0.8}^{+1.0}$& $3.4_{-0.9}^{+2.2}$ \\
$\chi^2/$dof & 7.0/6 &  11.9/9 & 18.4/12 & 1.9/5 \\
Flux ($10^{-15}$ erg cm$^{-2}$ s$^{-1}$) & 5.7 & 8.1 & 9.7 & 2.4 \\
Luminosity ($10^{38}$ erg s$^{-1}$) & 4.3 & 5.6 & 7.6 & 1.7 \\ 
\hline\hline 
\end{tabular} 

\end{center} 

\tablecomments{The $N_H$ values are in addition to the Galactic one.
Fluxes are observed values, luminosities are corrected for 
absorption, and are
calculated for the 0.3--8 keV band. Errors give the 68\% confidence
interval for one interesting parameter.}
\end{table*}

\clearpage

\begin{deluxetable}{lcccccccc} 
\rotate
\tablecaption{Spectral analysis of  the diffuse emission (projected regions).\label{denstemp}} 
\tabletypesize{\scriptsize}
\tablewidth{0pt}
\tablehead{
\colhead{Radius} &
\colhead{Net counts} &
\colhead{$\Gamma$} & 
\colhead{F(0.3--8 keV)} &
\colhead{L(0.3--8 keV)} &
\colhead{kT} & 
\colhead{F(0.3--8 keV)} &
\colhead{L(0.3--8 keV)} & 
\colhead{$\chi^2/$dof}
\\
\colhead{ } &
\colhead{ } &
\colhead{ } &
\colhead{($10^{-15}$ erg cm$^{-2}$ s$^{-1}$)} &
\colhead{($10^{38}$ erg s$^{-1}$)} &
\colhead{(keV)} & 
\colhead{(10$^{-15}$ erg cm$^{-2}$ s$^{-1}$)} &
\colhead{($10^{38}$ erg s$^{-1}$)} &
\colhead{ } 
}
\startdata
5$^{\prime\prime}$  & 166$\pm 13$ & $1.77_{-0.17}^{+0.17}$&
8.4$\pm 0.8$       & 6.7$\pm 0.7$ & --       & --  & --            & 8.6/9   \\
                    &             &  $1.61_{-0.21}^{+0.23}$& 8.6$\pm 1.2$      & 6.7$\pm 1.0$ & $0.21^{+0.16}_{-0.20}$  & 
0.37$_{-0.18}^{+13.6}$  & 0.44$_{-0.22}^{+16.2}$       & 7.4/7 \\
10$^{\prime\prime}$ & 175$\pm 16$ & $1.81_{-0.13}^{+0.13}$&
9.8$\pm 0.7$ & $7.9\pm 0.6$ & --       & --      & --   & 25.0/19 \\
      &          &  $1.57_{-0.15}^{+0.12}$& 10.0$\pm 1.4$ & $7.8\pm 1.1 $ & $0.25^{+0.08}_{-0.06}$ & 0.69$^{+0.35}_{-0.32}$ & $0.75^{+0.38}_{-0.35}$ & 21.9/17     \\
30$^{\prime\prime}$ & 581$\pm 48$ & $1.74_{-0.14}^{+0.14}$&
17.9$\pm 1.3$ & $14.3\pm 1.10$ &
--       & --      & --   & 78.2/64       \\
      &           & $1.59_{-0.17}^{+0.09}$& 16.9$_{-2.3}^{+3.1}$ & $13.0^{+2.40}_{-1.70}$ & $0.59^{+0.18}_{-0.18}$ & 1.06$^{+0.42}_{-0.71}$
 & $0.96^{+0.38}_{-0.64}$ &  71.9/62  \\
\enddata
\tablecomments{
$N_H$ is fixed at the Galactic value of $6.2\times 10^{20}$ 
cm$^{-2}$ (Tab.~\ref{mainlog}); in the $apec$ spectral model, the abundance is
fixed at the solar value.\\
Fluxes are observed, luminosities are intrinsic.\\
Errors give the 68\% confidence range for one interesting parameter.
}
\end{deluxetable}

\clearpage

\clearpage

\clearpage

\clearpage

\clearpage

\clearpage

\clearpage

\clearpage

\clearpage

\clearpage

\clearpage

\clearpage

\clearpage

\end{document}